\documentclass[pre,twocolumn,showpacs]{revtex4}

\usepackage{graphicx,epsfig,verbatim,enumerate}
\usepackage{amssymb}
\usepackage{ifthen}

\newcommand{\thnote}[1]{}
\newcommand{\footcite}[1]{~\cite{#1}}

\newcommand{\www}[1]{\url{#1}}
\newcommand{\req}[1]{(\ref{#1})}

\newcommand{\Om}{\Omega}
\newcommand{\om}{\omega}

\newcommand{\dee}[1]{\mbox{d}#1}

\newcommand{\avg}[1]{\left\langle#1\right\rangle}
\newcommand{\tavg}[1]{\langle#1\rangle}

\newcommand{\okell}{{l^{\mbox{\scriptsize{(s)}}}}}

\newcommand{\okellsom}{l_{\om}^{\mbox{\scriptsize{\, (s)}}}}
\newcommand{\okellsombar}{\bar{l}_{\om}^{\mbox{\scriptsize{\, (s)}}}}

\begin{document}

\title{
Geometry of River Networks II:\\
Distributions of Component Size and Number
}

\author{
  \firstname{Peter Sheridan}
  \surname{Dodds}
  }
\thanks{Author to whom correspondence should be addressed}
\email{dodds@segovia.mit.edu}
\homepage{http://segovia.mit.edu/}
\affiliation{Department of 
Mathematics and Department of Earth, 
Atmospheric and Planetary Sciences,
Massachusetts Institute of Technology,
Cambridge, MA 02139.}

\author{
  \firstname{Daniel H.}
  \surname{Rothman}
  }
\email{dan@segovia.mit.edu}
\affiliation{Department  of Earth, 
Atmospheric and Planetary Sciences,
Massachusetts Institute of Technology, 
Cambridge, MA 02139.}

\date{\today}

\begin{abstract}
The structure of a river network may be seen as a discrete set of
nested sub-networks built out of individual stream segments.  These
network components are assigned an integral stream order via a
hierarchical and discrete ordering method.  Exponential relationships,
known as Horton's laws,  between stream order and ensemble-averaged
quantities pertaining to network components are observed.  We extend
these observations to incorporate fluctuations and all higher moments
by developing functional relationships between distributions.  The
relationships determined are drawn from a combination of  theoretical
analysis,  analysis of real river networks including the Mississippi,
Amazon and Nile, and numerical simulations on a model of directed,
random networks.  Underlying distributions of stream segment lengths
are identified as exponential.  Combinations of these distributions
form single-humped distributions with exponential tails, the sums of
which are in turn shown to  give power law distributions of stream
lengths.  Distributions of basin area and  stream segment frequency
are also addressed.  The calculations identify a  single length-scale
as a measure of size fluctuations in network components.  This article
is the second in a series of three addressing the geometry of river networks.
\end{abstract}


\maketitle

\section{Introduction}
\label{sec:horton.intro}
Branching networks are an important
category of all networks
with river networks being a paradigmatic example.
Probably as much as any other natural phenomena,
river networks are a rich source
of scaling laws~\cite{rodriguez-iturbe97,rinaldo98,dodds2000pa}.
Central quantities such as drainage basin area
and stream lengths are reported to closely obey
power-law statistics~\cite{rodriguez-iturbe97,rinaldo98,dodds2000pa,langbein47,hack57,abrahams84,maritan96a,dodds99pa}.
The origin of this scaling has been
attributed to a variety of mechanisms including,
among others: 
principles of optimality~\cite{rodriguez-iturbe97,sun94}, 
self-organized criticality~\cite{rinaldo93},
invasion percolation~\cite{stark91},
and random fluctuations~\cite{dodds2000pa,leopold62,scheidegger67,manna92}.
One of the difficulties in establishing any theory
is that the reported values of scaling exponents
show some variation~\cite{abrahams84,maritan96a,maritan96b}.

With this variation in mind,
we have in~\cite{dodds2000ua}
extensively examined Hack's law, the scaling relationship
between basin shape and stream length.
Such scaling laws are inherently 
broad-brushed in their descriptive content.
In an effort to further improve comparisons between
theory and data and, more importantly,
between networks themselves,
we consider here a generalization
of Horton's laws~\cite{horton45,schumm56a}.  
Defined fully in the following section,
Horton's laws describe how average values of
network parameters change with a certain discrete
renormalization of the network.
The introduction of these laws
by Horton\nocite{horton45}
may be seen as one of many examples
that presaged the theory of fractal geometry~\cite{mandelbrot83}.
In essence, they express the relative frequency
and size of network components such as stream segments
and drainage basins.  

Here, we extend Horton's laws to functional 
relationships between probability distributions
rather than simply average values.
The recent work of 
Peckham and Gupta
was the first to address this natural
generalization of Horton's laws~\cite{peckham99}.
Our work agrees with their
findings but goes further to
characterize the distributions and develop
theoretical links between the distributions
of several different parameters.
We also present empirical studies that reveal
underlying scaling functions
with a focus on fluctuations and
further consider deviations due to
finite-size effects.

We examine continent-scale networks:
the Mississippi, Amazon, Congo,
Nile and Kansas river basins.
As in~\cite{dodds2000ua}, we also examine
Scheidegger's model of directed, 
random networks~\cite{scheidegger67}.
Both real and model networks
provide important tests and motivations
for our generalizations of Horton's laws.

We begin with definitions of 
stream ordering and Horton's laws.
Thereafter, the paper is divided into
two main sections.
In Section~\ref{sec:horton.postform}, 
we first sketch the 
theoretical generalization of Horton's laws.
Estimates of the Horton ratios are carried
out in Section~\ref{sec:horton.hortonratios}
and these provide basic parameters of
the generalized laws.
Empirical evidence from
real continent-scale networks is then provided
along with data from Scheidegger's random network model
in Section~\ref{sec:horton.generalization}.
In Section~\ref{sec:horton.moments}
we derive the higher order moments for
stream length distributions and in
Section~\ref{sec:horton.devs},
we consider deviations from Horton's laws
for large basins.
In the Appendix~\ref{sec:horton.theory},
we expand on some of the connections 
outlined in Section~\ref{sec:horton.generalization}, 
presenting a number of mathematical
considerations on these generalized Horton distributions.

This paper is the second in a series of three
on the geometry of river networks.
In the first~\cite{dodds2000ua}
we address issues of scaling and universality
and provide further motivation for our
general investigation.
In the third article 
of the series~\cite{dodds2000uc}
we extend the work of the present paper
by examining how the
detailed architecture of river networks,
i.e., how network components fit together.

\section{Stream ordering and Horton's laws}
\label{sec:horton.defs}

Stream ordering was first introduced by Horton\nocite{horton45}
in an effort to quantify the features of 
river networks~\cite{horton45}.
The method was later improved by Strahler
to give the present technique of 
Horton-Strahler stream ordering~\cite{strahler57}.
Stream ordering is a method applicable
to any field where branching, hierarchical networks are important.
Indeed, much use of stream ordering has been made outside
of the context of river networks,
a good example being the study
of venous and arterial blood networks 
in biology~\cite{zamir83,fung90,kassab93a,kassab94a,kassab94b,turcotte98,aharinejad98,zamir99}.
We describe two conceptions of the method and then discuss
empirical laws defined with in the context of stream ordering.

\begin{figure}[tbh!]
  \begin{center}
    \epsfig{file=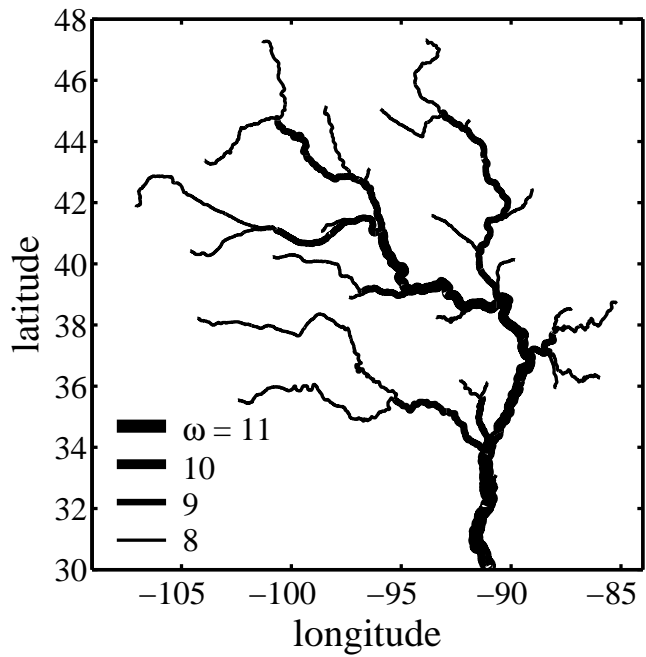,width=0.48\textwidth}
    \caption[Stream segment diagram of the Mississippi]{
      Stream segments for $\om=8$ up to $\om=\Om=11$ for
      the Mississippi River.  The spherical coordinates
      of latitude and longitude are used and the
      scale corresponds to roughly 2000 km along each axis.
      }
    \label{fig:horton.order_paths_mispi10}
  \end{center}
\end{figure}

A network's constituent stream segments are ordered
by an iterative pruning.  
An example of stream ordering for the Mississippi basin
is shown in Figure~\ref{fig:horton.order_paths_mispi10}.
A source stream is defined as a section of stream that runs
from a channel head to a junction with another stream
(for an arboreal analogy, think of the leaves of a tree).
These source streams are classified as the
first order stream segments of the network.  
Next, remove these source streams and 
identify the new source streams of the remaining network.  
These are the second order stream segments.  The process
is repeated until one stream segment is left of order $\Omega$.
The order of the network is then defined to be $\Omega$.

Once stream ordering on a network has been done,
a number of natural quantities arise.
These include $n_\om$, 
the number of basins (or equivalently stream segments)
for a given order $\om$;
$\tavg{l_\om}$, the average main stream length;
$\tavg{\okellsom}$, the average stream segment length;
$\tavg{a_\om}$, the average basin area; and the
variation in these numbers from order to order.  
Horton~\cite{horton45} and later 
Schumm~\cite{schumm56a} observed
that the following ratios
are generally independent of order $\om$:
\begin{equation}
  \frac{n_{\om}}{n_{\om+1}} = R_n, 
  \qquad
  \frac{\avg{l}_{\om+1}}{\avg{l}_{\om}} = R_l, 
  \qquad \mbox{and} \qquad
  \frac{\avg{a}_{\om+1}}{\avg{a}_{\om}} = R_a.
  \label{eq:horton.hortonratios}
\end{equation}
Since the main stream length averages
$\bar{l}_\om$ are combinations of
stream segment lengths
$\bar{l}_\om = \sum_{\nu=1}^\om \okellsombar$
we have that the Horton ratio for
stream segment lengths $R_\okell$ is equivalent to $R_l$.
Because our theory will start with the distributions
of $\okell$, we will generally use the ratio $R_\okell$ 
in place of $R_l$.

Horton's laws have remained something of a mystery
in geomorphology---the study of earth
surface processes and form---due to their 
apparent robustness and
hence perceived lack of physical (or geological) content.
However, statements that Horton's laws are 
``statistically inevitable''~\cite{kirchner93},
while possibly true, have not yet been based on
reasonable assumptions~\cite{dodds2000pa}.
Furthermore, many other scaling laws can be shown to
follow in part from Horton's laws~\cite{dodds99pa}.
Thus, Horton's laws being without content would imply
the same is true for those scaling laws that follow from them.
Other sufficient assumptions include uniform drainage density
(i.e., networks are space-filling) and self-affinity
of single channels.
The latter can be expressed 
as the relation~\cite{maritan96a,tarboton88,labarbera89,tarboton90}
\begin{equation}
  l \propto L_{\parallel}^{d},
  \label{eq:horton.selfaffinestreams}
\end{equation}
where $L_{\parallel}$ is the longitudinal diameter of a basin.
Scaling relations may be derived and the set
of relevant scaling exponents can be reduced to just two: $d$ as given
above and the ratio $\ln{R_\okell}/\ln{R_n}$~\cite{dodds99pa}.
Note that one obtains $R_a \equiv R_n$ so that only
the two Horton ratios $R_n$ and $R_\okell$ are independent.
Horton ratios are thus of central importance in the
full theory of scaling for river networks.

\section{Postulated form of Horton distributions}
\label{sec:horton.postform}

Horton's laws relate quantities which
are indexed by a discrete set of
numbers, namely the stream orders.
They also algebraically relate mean quantities
such as $\bar{a}_\om$.
Hence we may consider a generalization
to functional relationships between probability
distributions.  In other words, for stream lengths and
drainage areas we can explore
the relationships between
probability distributions defined for each order.

Furthermore, as we have noted,
Horton's laws can be used to derive
power laws of continuous variables
such as the probability distributions of 
drainage area $a$ and 
main stream length $l$~\cite{maritan96a,dodds99pa,devries94}:
\begin{equation}
  \label{eq:horton.Pa}
  P(a) \propto a^{-\tau}
  \qquad
  \mbox{and}
  \qquad
  P(l) \propto l^{-\gamma}.
\end{equation}
These derivations necessarily only give discrete
points of power laws.  
In other words,
the derivations give points as functions of the discrete stream 
order $\om$ and are uniformly spaced logarithmically
and we interpolate the power law from there.
The distributions
for stream lengths and areas
must therefore have structures that when combined
across orders produce smooth power laws.

For the example of the stream segment
length $\okellsom$, Horton's laws
state that the mean $\okellsombar$ grows by
a factor of $R_\okell$ with each integer step in order $\om$.
In considering $P(\okellsom,\om)$, the 
underlying probability distribution function for $\okellsom$,
we postulate that Horton's laws apply for every
moment of the distribution and not just the mean.
This generalization of Horton's laws may
be encapsulated in a statement about the
distribution $P(\okellsom,\om)$ as
\begin{equation}
  \label{eq:horton.ellwfreq}
  P(\okellsom,\om) = c_\okell(R_nR_\okell)^{-\om} F_\okell(\okellsom R_\okell^{-\om}).
\end{equation}
The factor of $(R_n)^{-\om}$ indicates the that 
$\int_{\okell=0}^\infty \dee{\okell} P(\okellsom,\om) \propto (R_n)^{-\om}$,
i.e., the frequency of stream segments of order $\om$ decays
according to Horton's law of stream number
given in equation~\req{eq:horton.hortonratios}.
Similarly, for $l_\om$, $a_\om$ and $n_{\Om,\om}$, we write
\begin{equation}
  \label{eq:horton.lwfreq}
  P(l_\om,\om) = c_l(R_nR_\okell)^{-\om} F_l(l_\om R_\okell^{-\om}),
\end{equation}
\begin{equation}
  \label{eq:horton.awfreq}
  P(a_\om,\om) = c_a(R_n^2)^{-\om} F_a(a_\om R_n^{-\om}),
\end{equation}
and
\begin{equation}
  \label{eq:horton.nwfreq}
  P(n_{\Om,\om}) = c_n(R_n)^{\Om-\om} F_n(n_{\Om,\om} R_n^{-\om}),
\end{equation}
where constants $c_\okell$, $c_l$, $c_a$ and $c_n$ are
appropriate normalizations.
We have used the subscripted versions of the lengths
and areas,  $\okellsom$, $l_\om$, and $a_\om$, to
reinforce that these parameters are for points
at the outlets of order $\om$ basins only.
The quantity $n_{\Om,\om}$ is the number of
streams of order $\om$ within a basin of order $\Om$.
This will help with some notational issues later on.
The form of the distribution functions $F_\okell$, $F_l$, $F_a$ and $F_n$
and their interrelationships
become the focus of our investigations.
Since scaling is inherent in each of these
postulated generalizations of Horton's laws,
we will often refer to these distribution functions
as \textit{scaling functions}.

We further postulate that distributions of stream
segment lengths are best approximated by exponential distributions.
Empirical evidence for this will be provided later on
in Section~\ref{sec:horton.generalization}.
The normalized scaling function $F_\okell(u)$ of
equation~\req{eq:horton.ellwfreq} then has the form
\begin{equation}
  \label{eq:horton.elldist}
  F_\okell(u) = \frac{1}{\xi} e^{-u/\xi} = F_\okell(u; \xi),
\end{equation}
where we have introduced a new
length scale $\xi$ and stated its appearance with
the notation $F_\okell(u; \xi)$.
The value of $\xi$ is potentially
network dependent.
As we will show, distributions of main stream
lengths, areas and stream number are all
dependent on $\xi$ and this is the only
additional parameter necessary for their description.
Note that $\xi$ is both the mean and
standard deviation of $F_\okell(u; \xi)$,
i.e., for exponential distributions,
fluctuations of a variable are on the order of its mean value.
We may therefore think of $\xi$ as a 
\textit{fluctuation length scale}.
Note that the presence of exponential distributions
indicates a randomness in the 
physical distribution
of streams themselves and this
is largely the topic of our third paper~\cite{dodds2000uc}.

Since main stream lengths are combinations of stream
segment lengths, i.e.\ 
$l_\om = \sum_{i=1}^{\om} \okellsom$, we have that
the distributions of main stream lengths of order $\om$
basins are approximated by convolutions of the
stream segment length distributions.
For this step,
it is more appropriate 
to use conditional probabilities such as $P(\okellsom| \om)$
where the basin order $\om$ is taken to be fixed.
We thus write
\begin{equation}
  \label{eq:horton.l-ellconv}
  P(l_\om|\om) =
  P(\okell_1|1) \ast P(\okell_2|2) \ast \cdots P(\okellsom|\om).
\end{equation}
where $\ast$ denotes convolution.
Details of the form obtained 
are given in Appendix~\ref{subsec:horton.mainstreamdist}.

The next step takes us to the power law distribution for
main stream lengths.  Summing over all stream orders
and integrating over $u=l_\om$ we have
\begin{equation}
  \label{eq:horton.l-lom}
  P(l) \simeq
\sum_{\om=1}^\infty \int_{u=l}^\infty \dee{u} P(u,\om),
\end{equation}
where we have returned to the joint probability
for this calculation.  
The integral over $u$ is replaced by a sum
when networks are considered on discrete lattices.
Note that the probability of finding a main stream
of length $l$ is independent of
any sort of stream ordering since
it is defined on an unordered network.
The details of this calculation
may be found in Appendix~\ref{subsec:horton.powerlawdists}
where it is shown that a power law $P(l) \propto l^{-\gamma}$
follows from the deduced form of the $P(l_\om,\om)$
with $\gamma = \ln{R_n}/\ln{R_\okell}$.

\section{Estimation of Horton ratios}
\label{sec:horton.hortonratios}

We now examine the usual Horton's laws
in order to estimate the Horton ratios.
These ratios are seen as intrinsic
parameters in the probability distribution 
functions given above in equations~\req{eq:horton.ellwfreq},
\req{eq:horton.lwfreq}, \req{eq:horton.awfreq} and~\req{eq:horton.nwfreq}.

\begin{figure}[tbp!]
  \begin{center}
   \ifthenelse{\boolean{@twocolumn}}
    {
    \begin{tabular}{c}
      \textbf(a) \\ 
      \epsfig{file=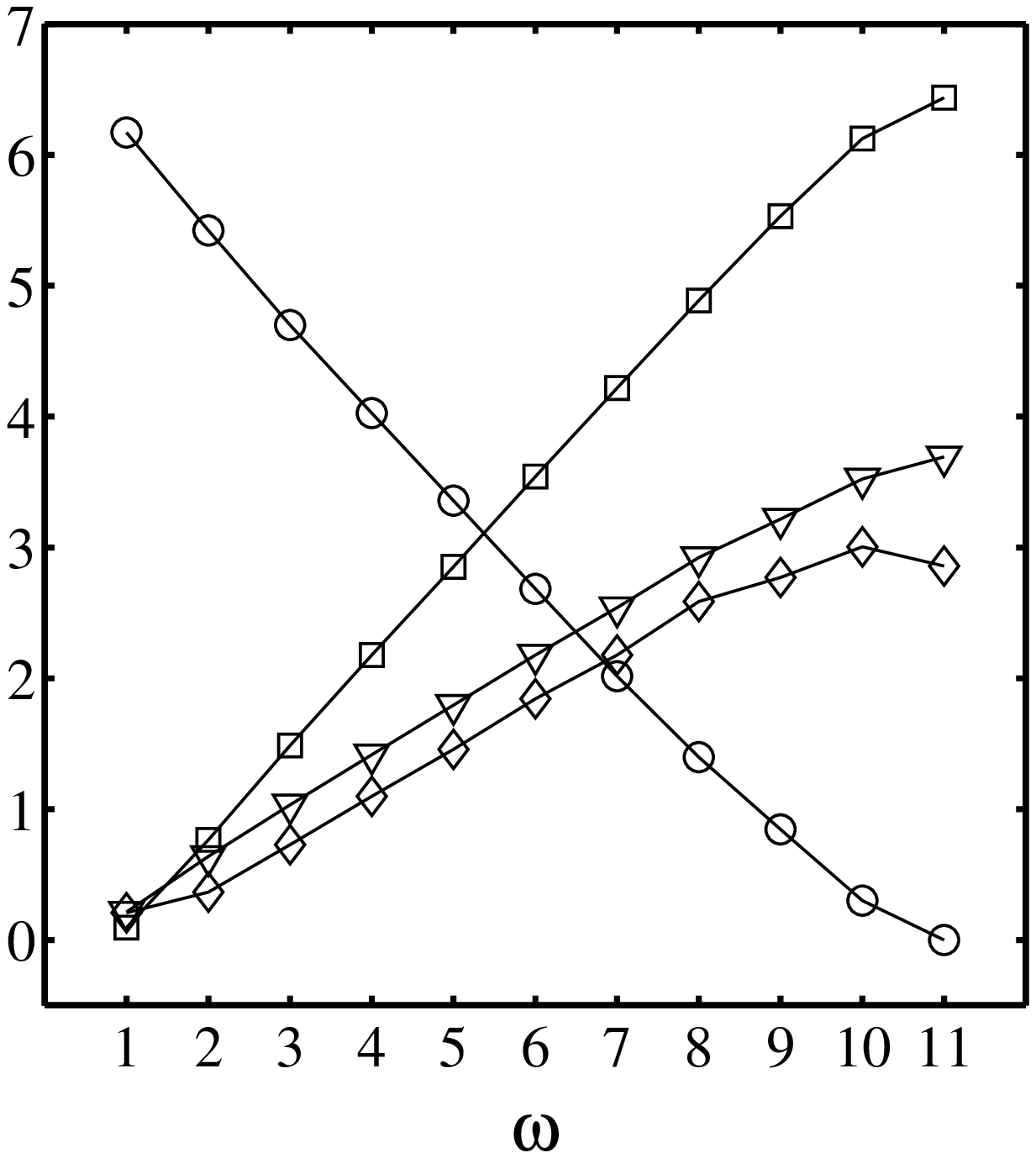,width=0.45\textwidth} \\
      \textbf(b) \\
      \epsfig{file=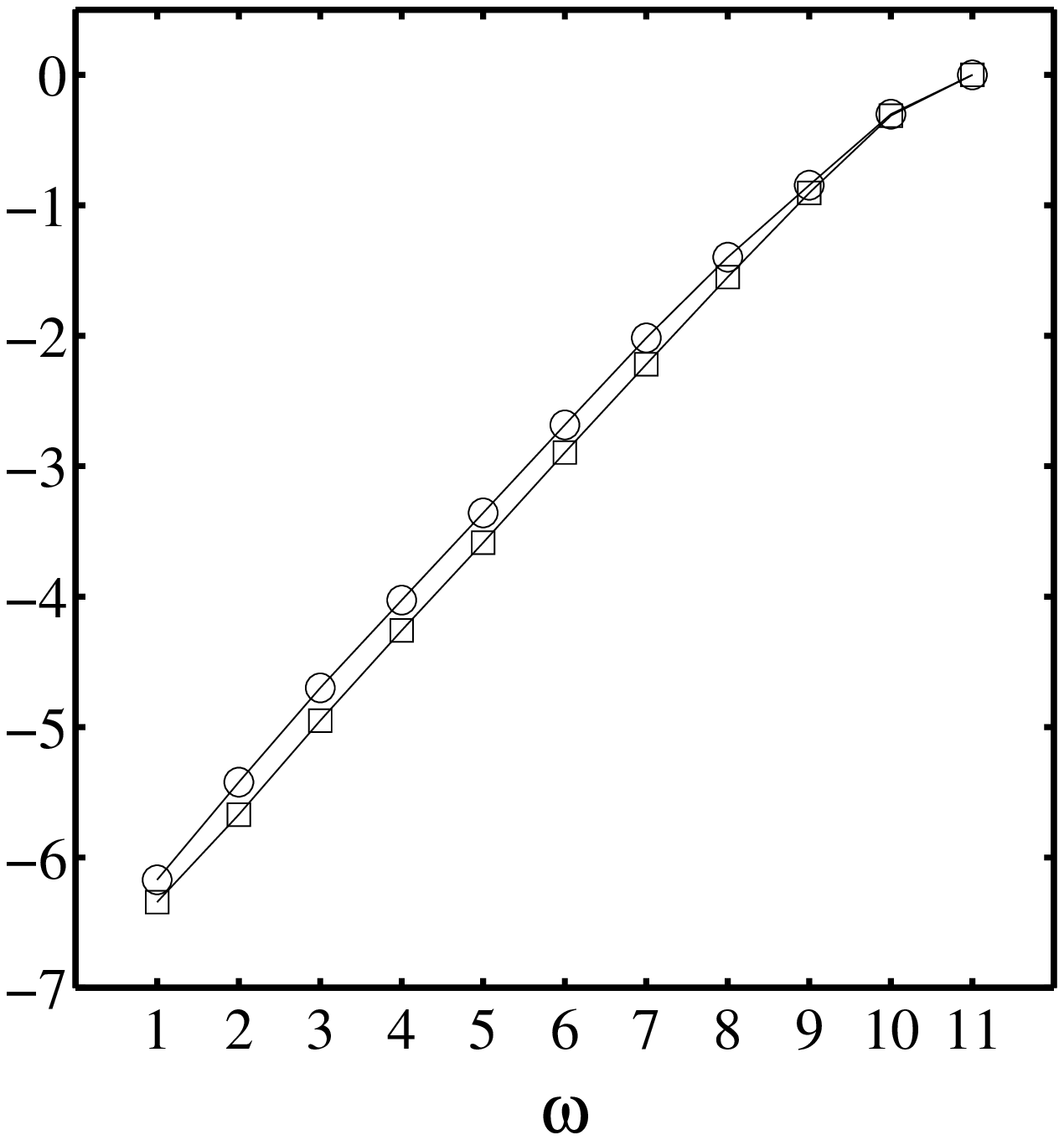,width=0.45\textwidth}
    \end{tabular}
    }
    {
    \begin{tabular}{cc}
      \textbf(a) & \textbf(b) \\
      \epsfig{file=fignalomega_mispi10_noname.ps,width=0.48\textwidth} & 
      \epsfig{file=fignflipaomega_mispi10_noname.ps,width=0.48\textwidth}
    \end{tabular}
    }
    \caption[Horton's laws for the Mississippi river basin network]{
      Horton's laws for the order $\Om=11$ 
      Mississippi river basin network.
      For (a), the ordinate axis is
      logarithmic (base 10) representing
      number for stream number $n_\om$ (circles),
      km$^2$ for area $\bar{a}_\om$ (squares),
      and km for both main stream length $\bar{l}_\om$ (triangles)
      and stream segment length $\okellsombar$ (diamonds).
      Note the good agreement between $\bar{l}_\om$
      and $\okellsombar$.
      In (b), the stream number data $n_\om$ (circles)
      has been inverted from that in (a), i.e., the 
      plot is of $n_{\om}^{-1}$.  
      This is compared with
      the dimensionless
      $\bar{a}_\om/\bar{a}_\Om$ (squares) showing good support
      for the prediction the slopes are equal, i.e.,
      $R_a \equiv R_n$.
      }
    \label{fig:horton.nalomega_mispi10}
  \end{center}
\end{figure}

\begin{table}[tb!]
  \begin{center}
    \begin{tabular}{ccccccc}
      $\omega$ range & $R_n$ & $R_a$ & $R_l$ & $R_{\okell}$ & $R_a/R_n$ & $R_l/R_{\okell}$ \\
      $[2,3]$ & 5.27 & 5.26 & 2.48 & 2.30 & 1.00 & 1.07 \\
      $[2,5]$ & 4.86 & 4.96 & 2.42 & 2.31 & 1.02 & 1.05 \\
      $[2,7]$ & 4.77 & 4.88 & 2.40 & 2.31 & 1.02 & 1.04 \\
      $[3,4]$ & 4.72 & 4.91 & 2.41 & 2.34 & 1.04 & 1.03 \\
      $[3,6]$ & 4.70 & 4.83 & 2.40 & 2.35 & 1.03 & 1.03 \\
      $[3,8]$ & 4.60 & 4.79 & 2.38 & 2.34 & 1.04 & 1.02 \\
      $[4,6]$ & 4.69 & 4.81 & 2.40 & 2.36 & 1.02 & 1.02 \\
      $[4,8]$ & 4.57 & 4.77 & 2.38 & 2.34 & 1.05 & 1.01 \\
      $[5,7]$ & 4.68 & 4.83 & 2.36 & 2.29 & 1.03 & 1.03 \\
      $[6,7]$ & 4.63 & 4.76 & 2.30 & 2.16 & 1.03 & 1.07 \\
      $[7,8]$ & 4.16 & 4.67 & 2.41 & 2.56 & 1.12 & 0.94 \\
      mean $\mu$ & 4.69 & 4.85 & 2.40 & 2.33 & 1.04 & 1.03 \\
      std dev $\sigma$ & 0.21 &  0.13 & 0.04 & 0.07 & 0.03 & 0.03\\
      $\sigma/\mu$ & 0.045 &  0.027 & 0.015 & 0.031 & 0.024 & 0.027\\
    \end{tabular}
    \caption[Horton ratios for the Mississippi]{
      Horton ratios for the Mississippi River\footcite{note:horton.mispitable}.
      For each range of orders $(\om_1,\om_2)$, estimates of the ratios are
      obtained via simple regression analysis. \thnote{(See Table~\ref{tab:app-data.mispi10orders} for the full range).}
      For each quantity, a mean $\mu$, standard deviation $\sigma$
      and normalized deviation $\sigma/\mu$ are calculated.
      All ranges with $2 \le \om_1 < \om_2 \le 8$ are used in
      these estimates but not all are shown
      The values obtained for $R_l$ are especially robust while
      some variation is observed for the estimates of $R_n$ and $R_a$.
      Good agreement is observed between the ratios $R_n$ and $R_a$
      and also between $R_l$ and $R_\okell$.
      \thnote{\input{note:horton.mispitable}}
      }
    \label{tab:horton.mispi10orders}
  \end{center}
\end{table}

\begin{table}[tb!]
  \begin{center}
    \begin{tabular}{ccccccc}
      $\omega$ range & $R_n$ & $R_a$ & $R_l$ & $R_{\okell}$ & $R_a/R_n$ & $R_l/R_{\okell}$ \\
      $[2,3]$ & 5.05 & 4.69 & 2.10 & 1.65 & 0.93 & 1.28 \\
      $[2,5]$ & 4.65 & 4.64 & 2.11 & 1.92 & 1.00 & 1.10 \\
      $[2,7]$ & 4.54 & 4.63 & 2.16 & 2.11 & 1.02 & 1.03 \\
      $[3,4]$ & 4.54 & 4.73 & 2.10 & 2.01 & 1.04 & 1.05 \\
      $[3,6]$ & 4.51 & 4.62 & 2.15 & 2.15 & 1.02 & 1.00 \\
      $[3,8]$ & 4.44 & 4.55 & 2.19 & 2.23 & 1.02 & 0.98 \\
      $[4,6]$ & 4.52 & 4.59 & 2.18 & 2.24 & 1.02 & 0.97 \\
      $[4,8]$ & 4.42 & 4.51 & 2.21 & 2.27 & 1.02 & 0.97 \\
      $[5,7]$ & 4.39 & 4.62 & 2.25 & 2.39 & 1.05 & 0.94 \\
      $[6,7]$ & 4.19 & 4.55 & 2.26 & 2.40 & 1.09 & 0.94 \\
      $[7,8]$ & 4.50 & 4.21 & 2.15 & 2.12 & 0.94 & 1.02 \\
      mean $\mu$ & 4.51 & 4.58 & 2.17 & 2.15 & 1.01 & 1.02 \\
      std dev $\sigma$ & 0.17 &  0.12 & 0.05 & 0.19 & 0.03 & 0.08\\
      $\sigma/\mu$ & 0.038 &  0.026 & 0.024 & 0.089 & 0.034 & 0.078\\
    \end{tabular}
    \caption[Horton ratios for the Amazon]{
      Horton ratios for the Amazon\footcite{note:horton.amazontable}.
      Details are as per Table~\ref{tab:horton.mispi10orders}. \thnote{(Again, not all data is shown but is recorded later in Table~\ref{tab:app-data.amazonorders}).} \thnote{\input{note:horton.amazontable}}
        }
      \label{tab:horton.amazonorders}
  \end{center}
\end{table}

\begin{table}[tb!]
  \begin{center}
    \begin{tabular}{ccccccc}
      $\omega$ range & $R_n$ & $R_a$ & $R_l$ & $R_{\okell}$ & $R_a/R_n$ & $R_l/R_{\okell}$ \\
      $[2,3]$ & 4.78 & 4.71 & 2.47 & 2.08 & 0.99 & 1.19 \\
      $[2,5]$ & 4.55 & 4.58 & 2.32 & 2.12 & 1.01 & 1.10 \\
      $[2,7]$ & 4.42 & 4.53 & 2.24 & 2.10 & 1.02 & 1.07 \\
      $[3,5]$ & 4.45 & 4.52 & 2.26 & 2.14 & 1.01 & 1.06 \\
      $[3,7]$ & 4.35 & 4.49 & 2.20 & 2.10 & 1.03 & 1.05 \\
      $[4,6]$ & 4.38 & 4.54 & 2.22 & 2.18 & 1.03 & 1.02 \\
      $[5,6]$ & 4.38 & 4.62 & 2.22 & 2.21 & 1.06 & 1.00 \\
      $[6,7]$ & 4.08 & 4.27 & 2.05 & 1.83 & 1.05 & 1.12 \\
      mean $\mu$ & 4.42 & 4.53 & 2.25 & 2.10 & 1.02 & 1.07 \\
      std dev $\sigma$ & 0.17 &  0.10 & 0.10 & 0.09 & 0.02 & 0.05\\
      $\sigma/\mu$ & 0.038 &  0.023 & 0.045 & 0.042 & 0.019 & 0.045\\
    \end{tabular}
    \caption[Horton ratios for the Nile]{
      Horton ratios for the Nile\footcite{note:horton.niletable}.
      Details are as per Table~\ref{tab:horton.mispi10orders}. \thnote{(See Table~\ref{tab:app-data.nileorders} for all data).}
      Here $2 \le \om_1 < \om_2 \le 7$. \thnote{\input{note:horton.niletable}}
      }
    \label{tab:horton.nileorders}
  \end{center}
\end{table}

Figure~\ref{fig:horton.nalomega_mispi10}(a)
shows the stream order averages of $\okell$, $l$, $a$ and $n$
for the Mississippi basin.  Deviations from
exponential trends of Horton's laws are evident and indicated by deviations
from straight lines on the semi-logarithmic axis.
Such deviations are to be expected for the smallest and
largest orders within a basin~\cite{dodds99pa,dodds2000uc}.
For the smallest orders, the scale of the grid
used becomes an issue
but even with infinite
resolution, the scaling of lengths, areas
and number for low orders
cannot all hold at the same time~\cite{dodds99pa}.
For large orders, the decrease in sample
space contributes to these fluctuations
since the number of samples of order $\om$ 
streams decays exponentially 
with order as $(R_n)^{\Om-\om}$.
Furthermore, correlations with overall basin shape
provide another source of deviations~\cite{dodds2000uc}.
Nevertheless, in our theoretical 
investigations below we will presume exact scaling.
Note also that the equivalence of $R_n$ and $R_a$ is supported
by Figure~\ref{fig:horton.nalomega_mispi10}(b)
where the stream numbers $n_w$ have been inverted for comparison.
Similar agreement is found for the Amazon and Nile
as shown in Tables~\ref{tab:horton.mispi10orders},
\ref{tab:horton.amazonorders},
and~\ref{tab:horton.nileorders} which we now discuss.

Table~\ref{tab:horton.mispi10orders} shows the
results of regression on the Mississippi data for various
ranges of stream orders for 
stream number, area and lengths.
Tables~\ref{tab:horton.amazonorders}
and~\ref{tab:horton.nileorders} show the same results
carried out for the Amazon and Nile.
Each table presents estimates of
the four ratios $R_n$, $R_a$, $R_l$ and
$R_\okell$.  Also included
are the comparisons $R_a/R_n$
and $R_l/R_\okell$, both of which we expect to
be close to unity.
For each quantity, we calculate the 
mean $\mu$, standard deviation $\sigma$
and normalized deviation $\sigma/\mu$.

Note the variation of exponents with choice
of order range.  
This is the largest source
of error in the calculation of the
Horton ratios.  
Therefore, rather than taking
a single range of stream orders for the regression,
we examine a collection of ranges.
Also, the deviations for high and low orders observed in
Figures~\ref{fig:horton.nalomega_mispi10}(a)
and~\ref{fig:horton.nalomega_mispi10}(b)
do of course affect measurements of the Horton ratios.
In all cases, we have avoided using data for
the smallest and largest orders.

For the three example networks given here, 
the statements $R_a \equiv R_n$ and 
$R_l \equiv R_\okell$ are well supported.
The majority of ranges give $R_n/R_a$ and
$R_l/R_\okell$ very close to unity.
The averages are also close to
one and are different from unity mostly by within 1.0 and
uniformly by within 1.5 standard deviations.

The normalized deviations, ie., $\sigma/\mu$,
for the four ratios are all below $0.05$.
No systematic ordering of the $\sigma/\mu$ is observed.
Of all the data, the values for $R_l$ in
the case of the Mississippi are the most
notably uniform having $\sigma/\mu = 0.015$.
Throughout there is a slight trend for
regression on lower orders to overestimate
and on higher orders to underestimate
the average ratios, while
reasonable consistency is found at intermediate orders.

Thus, overall the ranges chosen in the tables
give a reasonably even set of estimates
of the Horton ratios and
we will use these averages as our estimates of the ratios.

\begin{figure}[tbp!]
  \begin{center}
   \ifthenelse{\boolean{@twocolumn}}
    {
    \begin{tabular}{c}
      \textbf(a) \\ 
      \epsfig{file=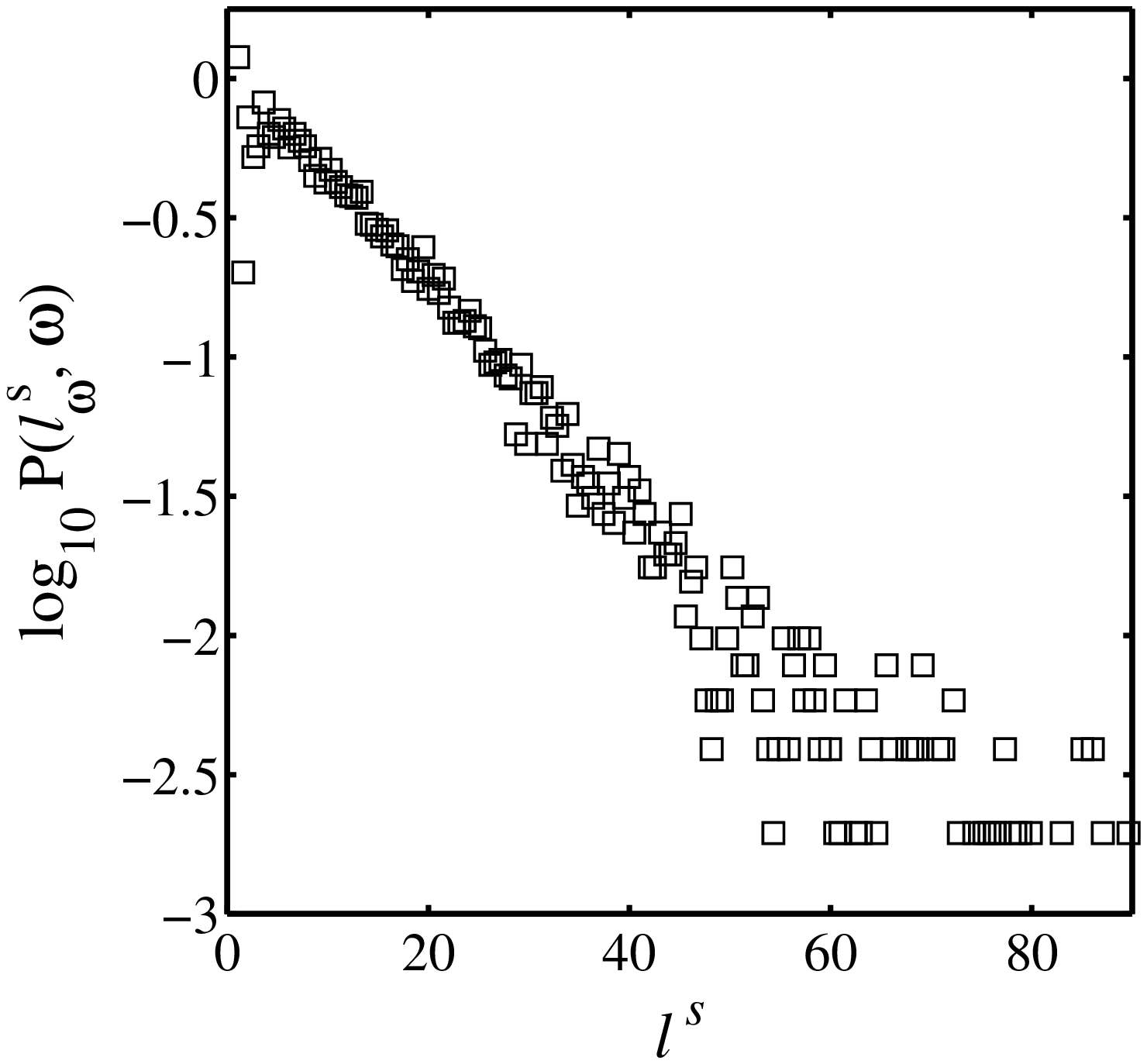,width=0.48\textwidth} \\
      \textbf(b) \\
      \epsfig{file=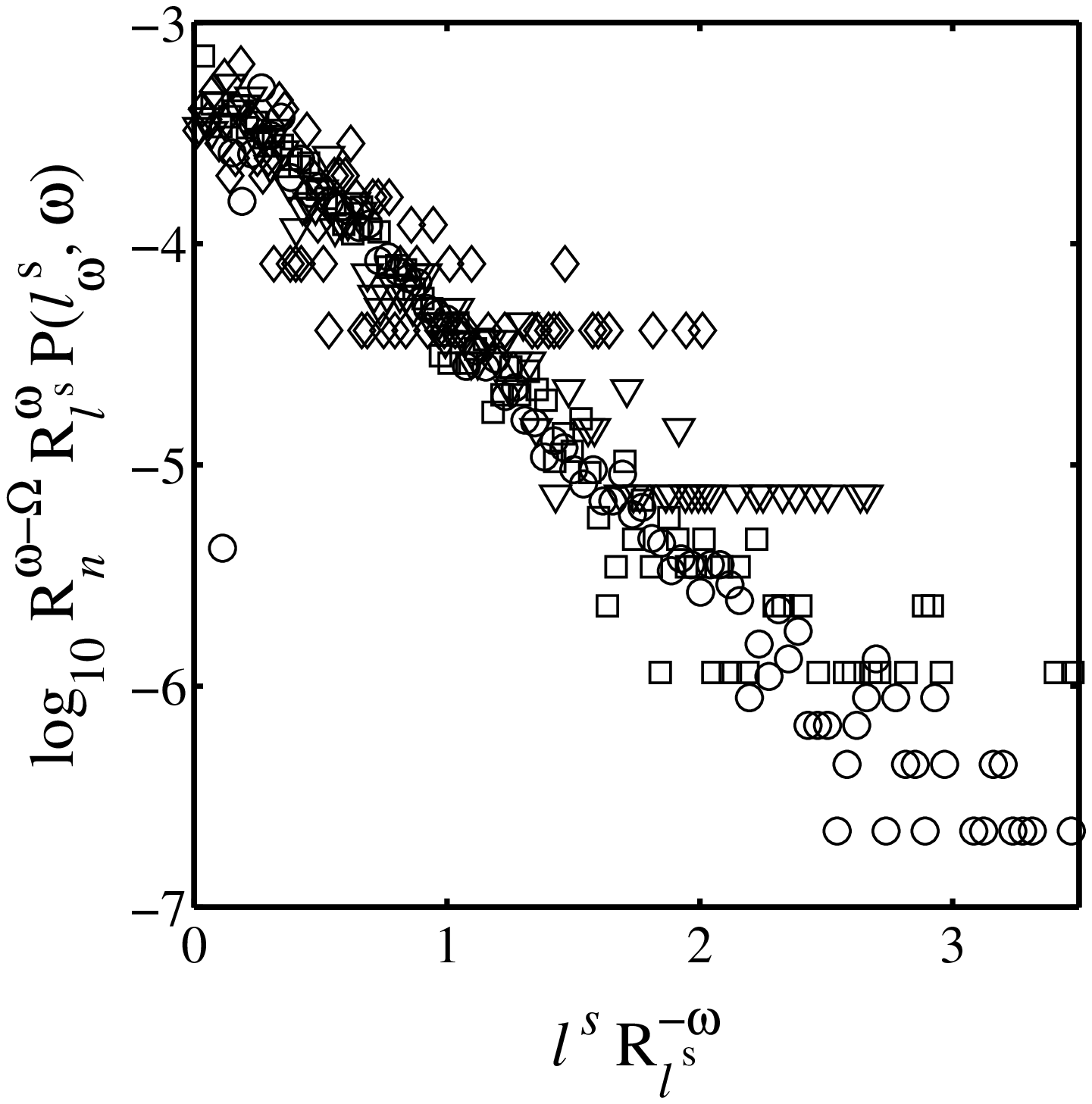,width=0.48\textwidth}
    \end{tabular}
    }
    {
    \begin{tabular}{cc}
      \textbf(a) & \textbf(b) \\
      \epsfig{file=figellw_collapse_mispi2_noname.ps,width=0.48\textwidth} &
      \epsfig{file=figellw_collapse_mispi2a_noname.ps,width=0.48\textwidth}
    \end{tabular}
    }
    \caption[Stream segment length distributions for the Mississippi]{
      Plot (a) shows an example distribution of  stream segment
      lengths, $P(\okellsom,\om)$, for the 
      Mississippi for order $\om=4$.  
      The lengths here are in kilometers.
      The semi-logarithmic axes indicate the distribution
      is well approximated by an exponential.
      The value of the length scale
      $\xi$ (see equation~\req{eq:horton.elldist}) is estimated 
      to be approximately 800 meters.
      Rescaled versions of the same stream segment length distributions
      for $\omega=3$ (circles), 
      $\omega=4$ (squares), 
      $\omega=5$ (triangles), 
      and
      $\omega=6$ (diamonds), are shown in (b).
      The rescaling is done according to 
      equation~\req{eq:horton.ellwfreq}.
      The values of the Horton ratios used
      are $R_n=4.69$ and $R_\okell = 2.33$
      as determined from Table~\ref{tab:horton.mispi10orders}
      }
    \label{fig:horton.ellw_collapse_mispi2}
  \end{center}
\end{figure}

\section{Empirical evidence for Horton distributions}
\label{sec:horton.generalization}

\subsection{Stream segment length distributions}
We now present Horton distributions for the Mississippi,
Amazon, and Nile river basins as well as the Scheidegger model.
Scheidegger networks may be thought of as collections of
random-walker streams and are fully defined
in~\cite{dodds2000ua} and extensively studied
in~\cite{dodds2000uc}.
The forms of all distributions are observed to be 
the same in the real data and in the model.

The first distribution
is shown in Figure~\ref{fig:horton.ellw_collapse_mispi2}(a).
This is the probability density function of $\okell_4$, 
fourth order stream segment
lengths, for the Mississippi River. 
Distributions for different orders
can be rescaled to show satisfactory agreement.
This is done using the postulated Horton distribution
of stream segment lengths given 
in equation~\req{eq:horton.ellwfreq}.
The rescaling is shown in 
Figure~\ref{fig:horton.ellw_collapse_mispi2}(b)
and is for orders $\om=3,\ldots,6$.
Note the effect
of the exponential decrease in number of
samples with order is evident for $\om=6$ since
$P(\okell_6)$ is considerably scattered.
Nevertheless, the figure shows the form of these distributions
to be most closely approximated by exponentials.
We observe similar exponential distributions for
the Amazon, the Nile and the Scheidegger model.
The fluctuation length scale $\xi$ is found to
be approximately $800$ meters for the Mississippi, $1600$ meters
for the Amazon and $1200$ meters for the Nile.

Since $\xi$ is based on the definition of stream ordering,
comparisons of $\xi$ are only sensible for networks that
are measured on topographies with the same resolution.
The above values of $\xi$ are approximate and our 
confidence in them would be improved
with higher resolution data.  Nevertheless they
do suggest that fluctuations in network structure
increase as we move from the Mississippi through
to the Nile and then the Amazon.

\subsection{Main stream segment length distributions}
\begin{figure}[tb!]
  \begin{center}
   \ifthenelse{\boolean{@twocolumn}}
    {
    \begin{tabular}{c}
      \textbf(a) \\ 
      \epsfig{file=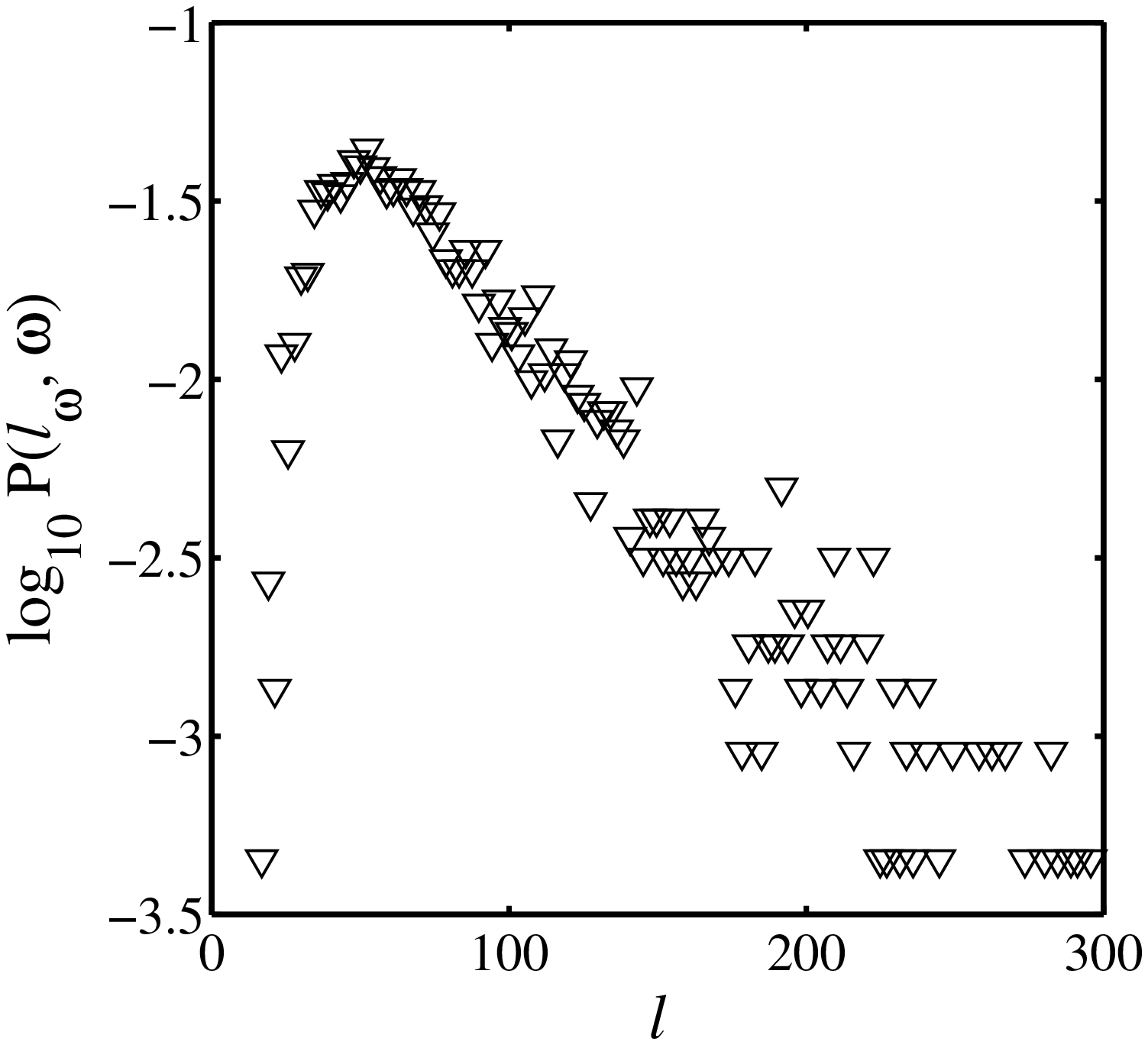,width=0.48\textwidth} \\
      \textbf(b) \\
      \epsfig{file=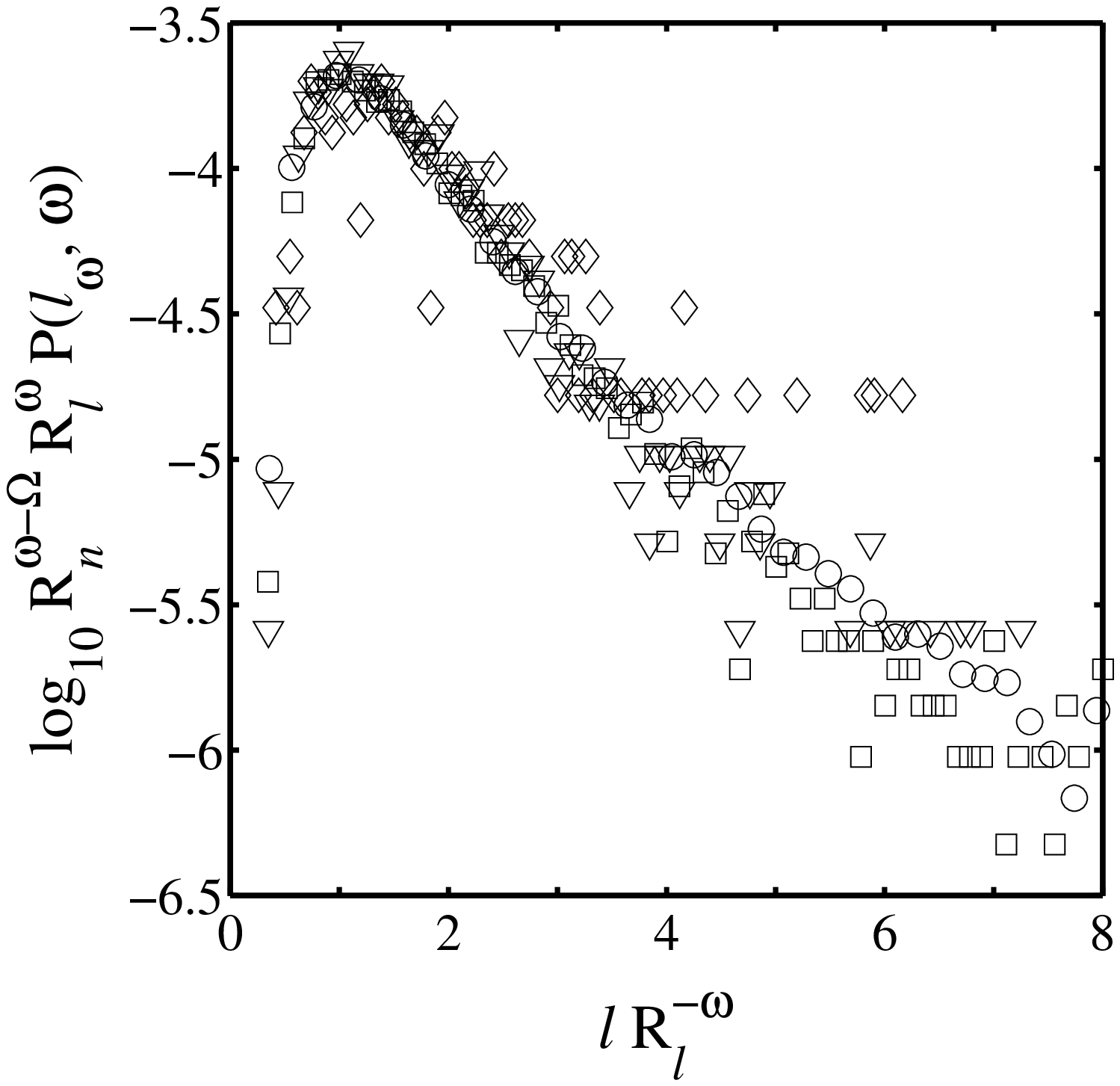,width=0.48\textwidth}
    \end{tabular}
    }
    {
    \begin{tabular}{cc}
      \textbf(a) & \textbf(b) \\
      \epsfig{file=figlw_collapse_amazon2b_noname.ps,width=0.48\textwidth} &
      \epsfig{file=figlw_collapse_amazon2a_noname.ps,width=0.48\textwidth}
    \end{tabular}
    }
    \caption[Main stream length distributions for the Amazon]{
      Plot (a) shows an example distribution
      for order $\om=5$ main stream
      lengths (measured in km) for the Amazon.
      The distribution is unimodal with
      what is a reasonable approximation of an exponential tail.
      In (b), distributions of main stream length for 
      for $\omega=3$ (circles), 
      $\omega=4$ (squares), 
      $\omega=5$ (triangles), 
      and
      $\omega=6$ (diamonds), are rescaled
      according to equation~\req{eq:horton.lwfreq}.
      The values of the Horton ratios used here
      are $R_n=4.51$ and $R_\okell = 2.17$,
      taken from Table~\ref{tab:horton.amazonorders}.
      }
    \label{fig:horton.lw_collapse_amazon2}
  \end{center}
\end{figure}

The distributions of $\om=4$ main stream lengths 
for the Amazon River is shown in
Figure~\ref{fig:horton.lw_collapse_amazon2}(a).
Since main stream lengths are sums of 
stream segment lengths, their distribution
has a single peak away from the origin.
However, these distributions will not tend 
towards a Gaussian because the individual
stream length distributions do not satisfy
the requirements of the central limit theorem~\cite{feller68I}.
This is because the moments of the stream segment
length distributions grow exponentially with stream order.
As the semi-logarithmic axes indicate, the tail
may be reasonably well (but not exactly) modeled by exponentials.
There is some variation in the distribution tails
from region to region.  
For example, corresponding distributions for the Mississippi data
do exhibit tails that are closer to exponentials.
However, for the present work where we
are attempting to characterize the basic forms
of the Horton distributions, we consider these
deviations to be of a higher order nature
and belonging to the realm of further research.

In accord with
equation~\req{eq:horton.lwfreq},
Figure~\ref{fig:horton.lw_collapse_amazon2}(b) shows
the rescaling of the main stream length distributions
for $\om=3,\ldots,6$.
The ratios used, $R_n=4.49$ and $R_l = 2.19 (\simeq R_\okell = 2.17)$
are taken from Table~\ref{tab:horton.amazonorders}.
Given the scatter of the distributions, it is unreasonable
to perform minimization techniques on the rescaled data itself
in order to estimate $R_n$ and $R_l$.  
This is best done by examining means, as we have
done, and higher order moments which we discuss below.
Furthermore, varying $R_n$ and $R_l$ from the above values
by, say, $\pm 0.05$ does not greatly distort the visual
quality of the ``data collapse.''

\begin{figure}[tbp!]
  \begin{center}
   \ifthenelse{\boolean{@twocolumn}}
    {
    \begin{tabular}{c}
      \textbf(a) \\ 
      \epsfig{file=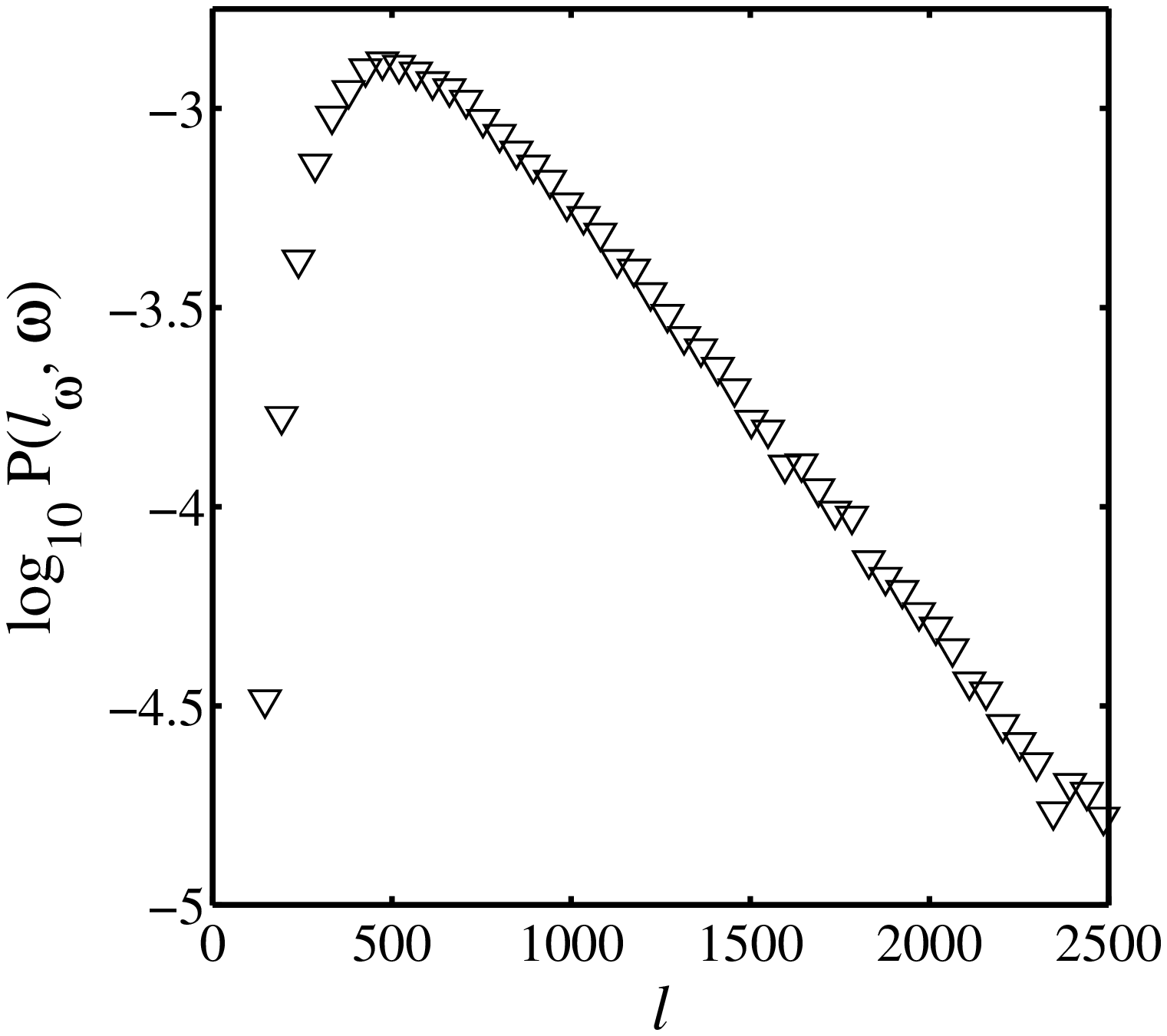,width=0.48\textwidth} \\
      \textbf(b) \\
      \epsfig{file=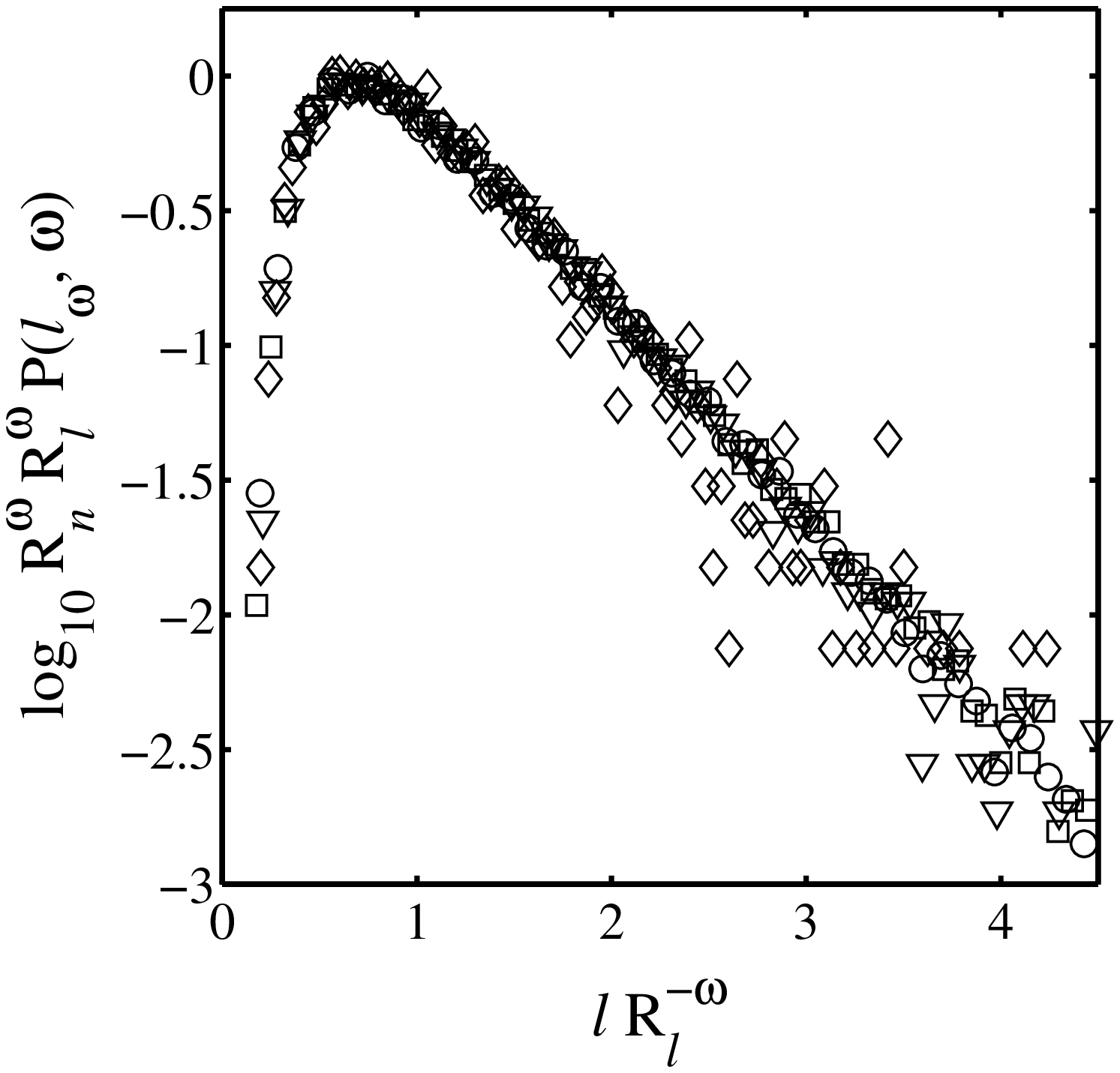,width=0.48\textwidth}
    \end{tabular}
    }
    {
    \begin{tabular}{cc}
      \textbf(a) & \textbf(b) \\
      \epsfig{file=figsche_lengthomega_6_noname.ps,width=0.48\textwidth} &
      \epsfig{file=figsche_lengthomega_collapse_noname.ps,width=0.48\textwidth}
    \end{tabular}
    }
    \caption[Main stream length distributions for the Scheidegger model]{
      Given in (a) is an example 
      distribution of order $\om=6$ 
      main stream lengths for the Scheidegger model.
      The same form is observed as for real networks
      such as the Amazon (Figure~\ref{fig:horton.lw_collapse_amazon2}).
      In the same way as Figure~\ref{fig:horton.lw_collapse_amazon2}(b),
      (b) show rescaled distributions of main stream length for 
      for $\omega=4$ (circles), 
      $\omega=5$ (squares), 
      $\omega=6$ (triangles), 
      and
      $\omega=7$ (diamonds).
      Note that in (b), distributions are not normalized
      with respect to a fixed basin order $\Om$ and
      hence the vertical offset is arbitrary.
      The values of the ratios used here are
      $R_n \simeq 5.20$ and $R_l \simeq 3.00$~\cite{dodds99pa}.
      }
    \label{fig:horton.figsche_lengthomega}
  \end{center}
\end{figure}

Similar results for the Scheidegger model
are shown in Figure~\ref{fig:horton.figsche_lengthomega}.
The Scheidegger model may be thought of as a network
defined on a triangular lattice where at each lattice 
site one of two directions is 
chosen as the stream path~\cite{dodds2000ua,dodds2000uc}.
Figure~\ref{fig:horton.figsche_lengthomega}(a) gives
a single example distribution for main stream lengths
of order $\om=6$ basins.  
The tail is exponential as per the real world data.
Figure~\ref{fig:horton.figsche_lengthomega}(b) shows
a collapse of main stream length distributions
for orders $\om=4$ through $7$.  
In contrast to the real data where
an overall basin order is fixed ($\Omega$),
there is no maximum basin order here.
The distributions in
Figure~\ref{fig:horton.figsche_lengthomega}(b)
have an arbitrary normalization meaning
the absolute values of the ordinate are also arbitrary.
Otherwise, this is the same collapse
as given in equation~\req{eq:horton.lwfreq}.
For the Scheidegger model, our simulations yield
$R_n \simeq 5.20$ and $R_\okell \simeq 3.00$~\cite{dodds99pa}.
For all distributions, we observe similar functional forms
for real networks and the Scheidegger model, the
only difference lying in parameters such
as the Horton ratios.

\subsection{Drainage area distributions}

\begin{figure}[tbp!]
  \begin{center}
   \ifthenelse{\boolean{@twocolumn}}
    {
    \begin{tabular}{c}
      \textbf(a) \\ 
      \epsfig{file=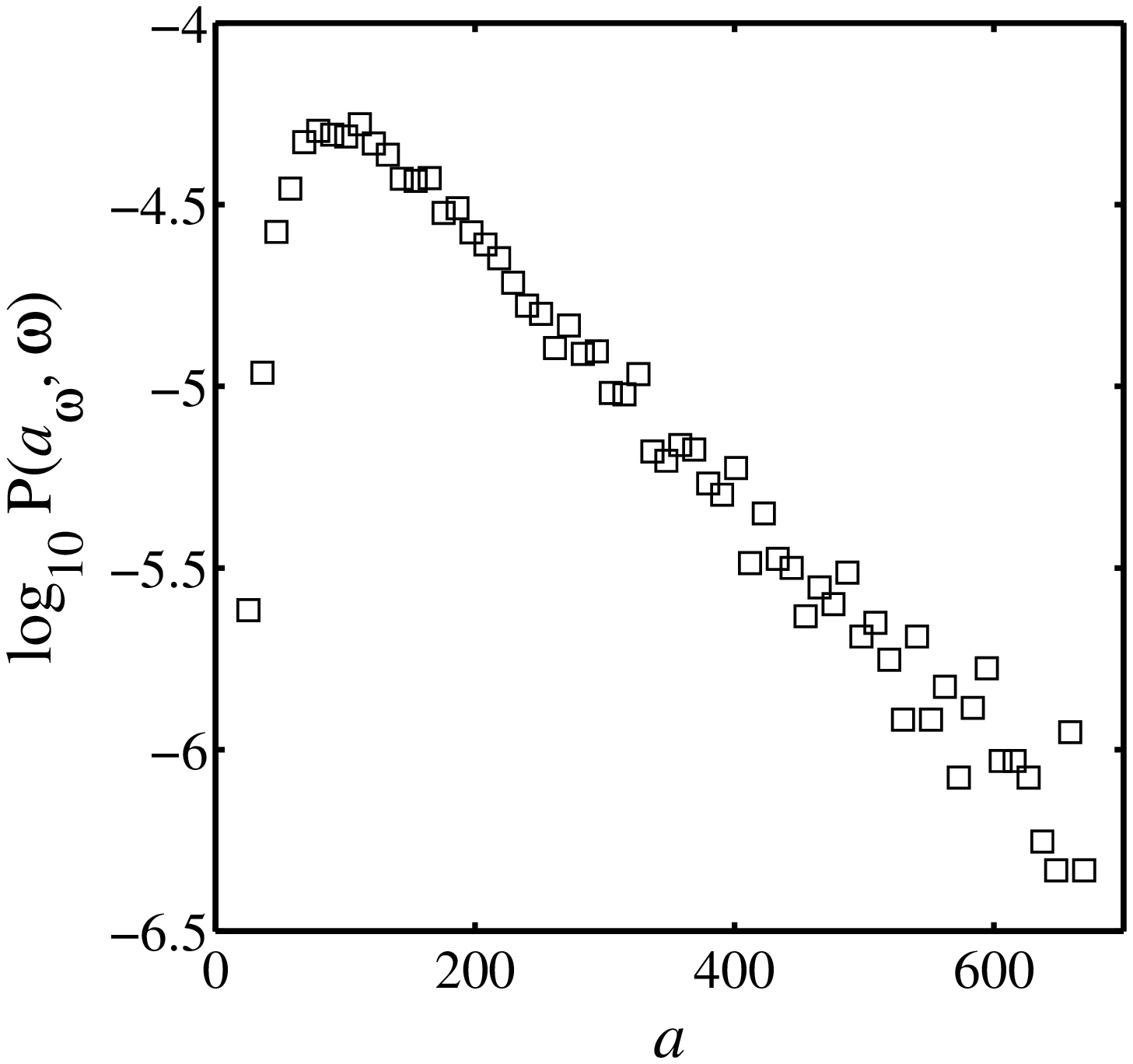,width=0.48\textwidth} \\
      \textbf(b) \\
      \epsfig{file=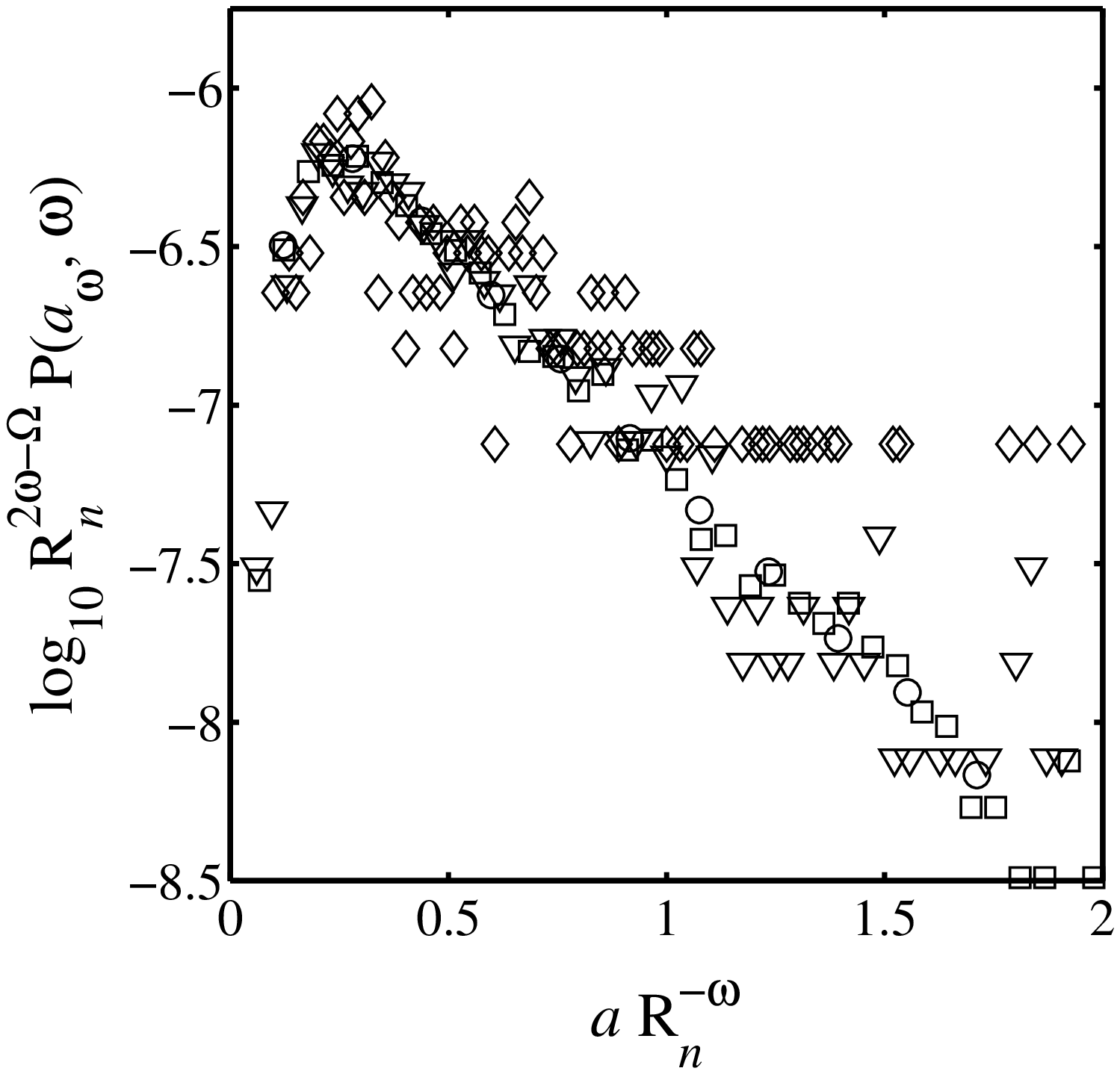,width=0.48\textwidth}
    \end{tabular}
    }
  {
    \begin{tabular}{cc}
      \textbf(a) & \textbf(b) \\
      \epsfig{file=figaw_collapse_nile2b_noname.ps,width=0.48\textwidth} &
      \epsfig{file=figaw_collapse_nile2a_noname.ps,width=0.48\textwidth}
    \end{tabular}
    }
    \caption[Drainage area distributions for the Nile]{
      The distribution of drainage areas for
      $\om=4$ sub-basins of the
      Nile are shown in (a).  
      All areas are measured in km$^2$.
      An exponential tail
      is observed as per the distributions
      of stream segment length (Figure~\ref{fig:horton.ellw_collapse_mispi2})
      and main stream length (Figure~\ref{fig:horton.lw_collapse_amazon2}).
      In (b), distributions of drainage area for 
      for $\omega=3$ (circles), 
      $\omega=4$ (squares), 
      $\omega=5$ (triangles), 
      and
      $\omega=6$ (diamonds), are rescaled
      according to equation~\req{eq:horton.awfreq}.
      The rescaling uses the estimate $R_n=4.42$ 
      found in Table~\ref{tab:horton.nileorders}.
      }
    \label{fig:horton.aw_collapse_nile2}
  \end{center}
\end{figure}

Figure~\ref{fig:horton.aw_collapse_nile2} shows
more Horton distributions, this
time for drainage area as calculated for the Nile river basin.
In Figure~\ref{fig:horton.aw_collapse_nile2},
an example distribution for $\om=4$ sub-basins is presented.
The distribution is similar in form to those of main stream lengths
of Figure~\ref{fig:horton.lw_collapse_amazon2},
again showing a reasonably clear exponential tail.
Rescaled drainage area distributions for
$\om=3,\ldots,6$ 
are presented in Figure~\ref{fig:horton.aw_collapse_nile2}(b).
The rescaling now follows equation~\req{eq:horton.awfreq}.
Note that if $R_n$ and $R_a$ were not equivalent,
the rescaling would be of the form
\begin{equation}
  \label{eq:horton.awfreq2}
  P(a_\om,\om) = c_a(R_nR_a)^{-\om} F_a(a_\om R_a^{-\om}).
\end{equation}
Since we have asserted that $R_n \equiv R_a$,
equation~\req{eq:horton.awfreq2} reduces
to equation~\req{eq:horton.awfreq}.
The Horton ratio used here is $R_n = 4.42$
which is in good agreement with $R_a = 4.53$,
the respective standard deviations being $0.17$ and $0.10$.
Both figures are taken from the data of Table~\ref{tab:horton.nileorders}.

\subsection{Summing distributions to form power laws}

\begin{figure}[tb!]
  \begin{center}
    \epsfig{file=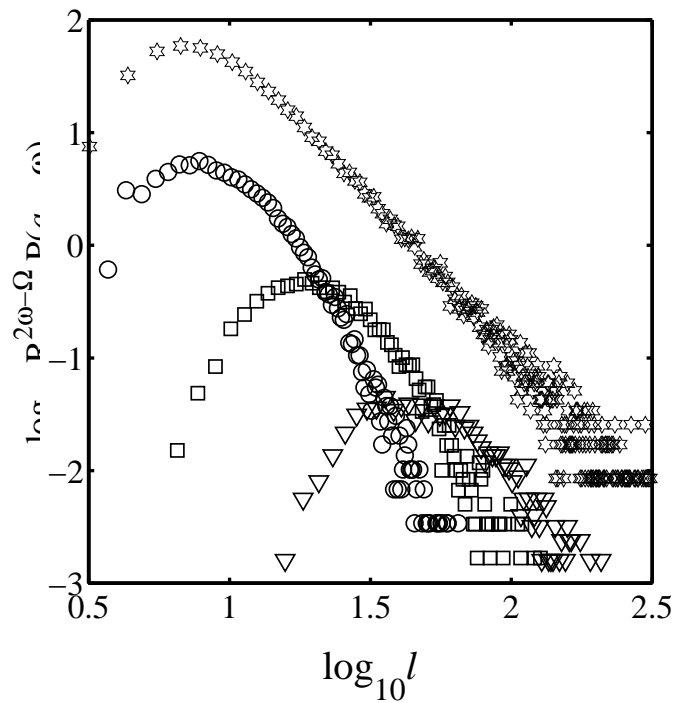,width=.48\textwidth}
    \caption[Summation of main stream length distributions for the Mississippi]{
      Summation of main stream length distributions for the Mississippi.
      Both axes are logarithmic,the
      unit of length is km and the vertical axis is probability
      density with units of km$^{-1}$.
      Distributions of $l_\om$ for 
      orders $\om=3$ (circles),
      $\om=4$ (squares),
      and $\om=5$ (triangles), are shown.
      As expected, the distributions
      sum together to give a power law tail (stars).
      The power law distribution 
      (which is vertically offset by an
      order of magnitude for clarity)
      is the summation of the distributions below 
      as well as the distribution for order $\om=6$ main
      stream lengths.  
      }
    \label{fig:lw_powerlawsum2_mispi}
  \end{center}
\end{figure}

As stated in Section~\ref{sec:horton.postform}, the Horton distributions 
of $a_\om$ and $l_\om$
must combine to form power law distributions
for $a$ and $l$ (see equations~\ref{eq:horton.Pa}
and \ref{eq:horton.l-lom}).
Figure~\ref{fig:lw_powerlawsum2_mispi}
provides empirical support for
this observation for the example main stream
lengths of the Mississippi network.
The distributions for $\om=3$, 4 and 5
main stream lengths are individually shown.
Their combination together with the distribution
of $l_6$ gives the reasonable approximation
of a power law as shown.
The area distributions combine in the same way.
Note that the distributions do not
greatly overlap.  
Each point of the
power law is therefore the addition of 
significant contributions from only two 
or three of the separate distributions.
The challenge here then is to understand
how rescaled versions of $F_l$,
being the basic form of the $P(l_\om,\om)$,
fit together in such a clean fashion.
The details of this connection are
established in Appendix~\ref{subsec:horton.powerlawdists}.

\subsection{Connecting distributions of number and area}
In considering the generalized Horton distributions
for number and area, we observe two main points:
a calculation in the vein of what we are able to do
for main stream lengths is difficult; 
and, the Horton distributions for 
area and number are equivalent.

In principle, Horton area distributions may be derived
from stream segment length distributions.  
This follows from an assumption of statistically
uniform drainage density which means that the
typical drainage area drained per unit length of any stream
is invariant in space.  Apart from the possibility of
changing with space which we will preclude by assumption, 
drainage density does naturally 
fluctuate as well~\cite{dodds2000ua}.
Thus, we can write $a \simeq \rho \sum_\om \okellsom$
where the sum is over all orders and all stream segments
and $\rho$ is the average drainage density.

However, we
need to know for an example basin,
how many instances of each stream segment occur as
a function of order.  
For example, the number of first
order streams in an order $\Om$ basins is $n_{\Om,1}$.
Given the distribution of this number, 
we can then calculate 
the distribution of the total contribution of drainage
area due to first order streams.  
But the
distributions of $n_{\Om,\om}$ are not independent
so we cannot proceed in this direction.

We could potentially use the typical number of order $\om$ streams,
$(R_n)^{\Om-\om}$.  Then the distribution of total area drained
due to order $\om$ streams would approach Gaussian because
the individual distribution are identical and the central
limit theorem would apply.  However, because the fluctuations in
total number of stream segments are so great, we 
lose too much information with this approach.  
Indeed,
the distribution of area drained by order $\om$ stream segments
in a basin reflects variations in their number rather
than length.  Again, we meet up with the problem of the
numbers of distinct orders of stream segment lengths being
dependent.

One final way would be to use Tokunaga's 
law~\cite{dodds99pa,tokunaga66,tokunaga78,tokunaga84,peckham95}.
Tokunaga's law states that the number of 
order $\nu$ 
side branches along an (absorbing) 
stream segment of order $\mu$ is given
by 
\begin{equation}
  \label{eq:horton.tokslaw}
  T_{k} = T_1 (R_\okell)^{k+1}.
\end{equation}
where $k=\mu-\nu$.  
The parameter $T_1$ is
the average number of side streams having order
$\nu=\mu-1$ for every order $\mu$ absorbing stream.
This gives a picture of how a network fits together
and may be seen to be equivalent to Horton's laws~\cite{dodds99pa}.
Now, even though we also
understand the distributions 
underlying Tokunaga's law~\cite{dodds2000ua},
similar technical problems arise.
On descending into a network, we find the number 
of stream segments at each level to be
dependent on all of the above.

\begin{figure}[tb!]
  \begin{center}
    \epsfig{file=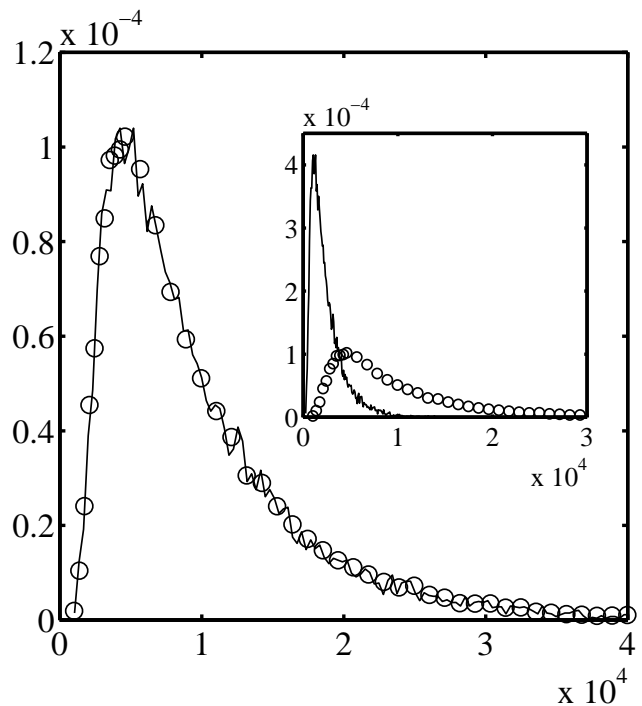,width=0.48\textwidth}
    \caption[Number and area distributions for the Scheidegger model]{
      Comparison of number and area distributions for the
      Scheidegger model.  
      Area is in terms of lattice units.
      In the inset plot,
      the raw distributions shown 
      are $P(a_6|6)$ (circles) and $P(n_{6,1}|6)$ (continuous line).
      The latter is the probability of finding $n_{6,1}$
      source streams in an order $\om=6$ basin.
      In the main plot, the number distribution has been
      rescaled to be $1/4P(n_{6,1}|6)$ as a function
      of $4n_{6,1}$ and the area distribution is unrescaled
      (the symbols are the same as for the inset plot).
      For the Scheidegger model, source streams occur at any site
      with probability of 1/4, hence the rescaling by a factor of four.
      }
    \label{fig:horton.figsche_number_area}
  \end{center}
\end{figure}

Nevertheless, we can understand the relationship between
the distributions for area and number.
What follows is a generalization of the
finding that $R_n \equiv R_a$.
The postulated forms for these distributions
were given in equations~\req{eq:horton.awfreq}
and~\req{eq:horton.nwfreq}.  
Consider $n_{\Om,1}$,
the number of first order streams in an
order $\Om$ basin.
Assuming that, on average, first order
streams are distributed evenly throughout
a network, then this number is simply
proportional to $a_\Om$.
As an example, 
Figure~\ref{fig:horton.figsche_number_area}
shows data obtained for the Scheidegger model.
For the Scheidegger model, first order streams
are initiated with a $1/4$ probability when
the flow at the two upstream sites is randomly directed
away, each with probability $1/2$.
Thus, for an area $a_\Om$, we expect 
and find $n_{\Om,\om}=a_{\Om}/4$.

For higher internal orders, we can apply
a simple renormalization.  Assuming
a system with exact scaling, the number
of streams $n_{\Om,\om}$ is statistically
equivalent to $n_{\Om-\om+1,1}$.
Since the latter is proportional to 
$a_{\Om-\om+1}$ we have that
\begin{equation}
  \label{eq:horton.n_a}
  n_{\Om,\om} \simeq \rho_\om a_{\Om-\om+1}
\end{equation}
where the constant of proportionality
is the density of order $\om$ streams, 
Clearly, this equivalence improves 
as number increases, i.e., the difference
$\Om-\om$ increases.

While we do not have exact forms for the
area or number distributions, we note
that they are similar to the main stream
length distributions.  
Since source streams are linear basins
with the width of a grid cell, the distribution
of $a_1$ is the same as the distribution
of $l_1$ and $\okell_1$, a pure exponential.
Hence, $n_{\Om,\Om-1}$ is also an exponential.
For increasing $\om$, the distribution of $a_\om$ 
becomes single peaked with an exponential tail,
qualitatively the same as the main stream length
distributions.

\begin{figure}[tb!]
  \begin{center}
    \epsfig{file=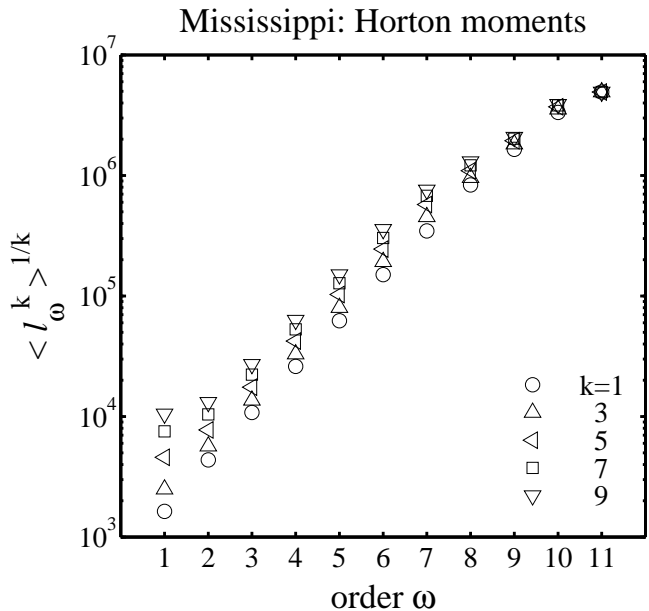,width=0.48\textwidth}
    \caption[Moment comparison for main stream lengths for the Mississippi]{
      A comparison of moments calculated for 
      main stream length distributions
      for the  Mississippi River.
      }
    \label{fig:horton.figmoments_l_mispi}
  \end{center}
\end{figure}

\section{Higher order moments}
\label{sec:horton.moments}

Finally, we discuss the higher order moments
for the generalized Horton distributions.
Figure~\ref{fig:horton.figmoments_l_mispi}
presents moments for distributions 
of main stream lengths for the case of the Mississippi.
These moments are calculated directly from 
the main stream length distributions.  
A regular logarithmic spacing is 
apparent in moments for
orders ranging from $3$ to $7$.

To see whether or not this is expected, 
we detail a few small calculations concerning
moments starting from the exponential form
of stream segment lengths given
in equation~\req{eq:horton.elldist}.
As noted previously,
for an exponential distribution,
$F_\okell(u) = \xi^{-1} e^{-u/\xi}$,
the mean is simply $\tavg{u} = \xi$.
In general, the $q$th moment
of an exponential distribution is
\begin{eqnarray}
  \label{eq:horton.highmom_exp}
  \avg{u^q} & = & \int_{u=0}^{\infty} \frac{u^q}{\xi} e^{-u/\xi} \dee{u} \nonumber \\
  & = & \xi^q \int_{x=0}^{\infty} x^q e^{-x} \dee{x} =  q! \xi^q.
\end{eqnarray}
Assuming
scaling holds exactly for across all orders,
the above is precisely $\tavg{(\okell_1)^q}$.
Note that $\tavg{(\okell_1)^q} = q!\tavg{\okell_1}^q$.
Since the characteristic length of
order $\om$ streams is $(R_\okell)^{\om-1}$,
we therefore have
\begin{equation}
  \label{eq:horton.highmom_exp_omega}
  \avg{(\okellsom)^q}
   = q! \xi^q (R_\okell)^{(\om-1)q} = q!\avg{\okellsom}^q.
\end{equation}

Since main stream lengths are sums of stream
segment lengths, so are their respective moments.
Hence,
\begin{eqnarray}
  \label{eq:horton.highmom_l_omega}
  \avg{(l_\om)^q} & = &
  \sum_{k=1}^{\om} \avg{(\okell_k)^q}, \nonumber \\
  & = & \sum_{k=1}^{\om} q! \xi^q (R_\okell)^{(k-1)q}, \nonumber \\
  & = & q! \xi^q \sum_{k=1}^{\om} (R_\okell)^{(k-1)q}, \nonumber \\
  & = & q! \xi^q \frac{(R_\okell)^{q\om}-1}{R_\okell - 1}.
\end{eqnarray}
We can now determine the log-space separation of 
moments of main stream length.
Using Stirling's approximation~\cite{gradshteyn65}
that $\ln{n!} \sim (n+1/2)\ln{n} -n$
we have
\begin{equation}
  \label{eq:horton.logsp_mom}
  \ln{\avg{(l_\om)^q}} 
  \sim q \left[\xi + (R_\okell)^{\om} + \ln{q} \right]+ C,
\end{equation}
where $C$ is a constant.  
The $\ln{q}$ term inside
the square brackets in equation~\ref{eq:horton.logsp_mom} creates
small deviations from linearity for $1 \le \om \le 15$.
Thus, in agreement with Figure~\ref{fig:horton.figmoments_l_mispi},
we expect approximately linear
growth of moments in log-space.

\section{Limitations on the predictive power of Horton's laws}
\label{sec:horton.devs}

In this last section, we briefly examine deviations from 
scaling within this generalized picture of Horton's laws.
The basic question is
given an approximate scaling for quantities measured
at intermediate stream orders, what can we say about
the features of the overall basin?

\begin{figure}[htb!]
  \begin{center}
    \epsfig{file=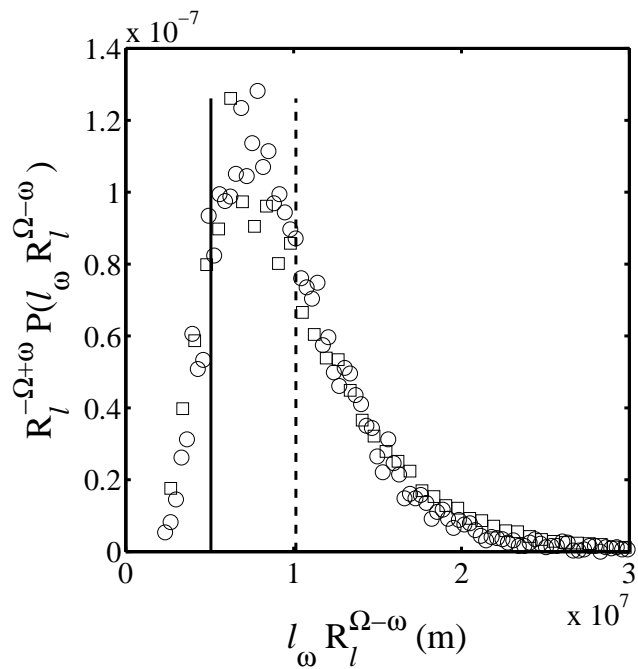,width=0.48\textwidth}
    \caption[Testing prediction of main stream length for the Congo]{
      Comparing the generalized Horton length
      distribution
      rescaled to the level of order 
      $\Om=11$ basins with the Congo River itself.
      The two distributions are for orders $\om=3$ (squares)
      and $\om=4$ (circles) stream lengths
      and the Horton ratio is estimated to
      be $R_l = 2.39$~\cite{dodds2000ub}.
      The dashed line
      represents the mean of these scaled up distributions
      while the solid line marks $\bar{l}_{11}$, the measured
      length of the Congo at a 1000 meter resolution.
      The actual length is within a standard deviation of
      the mean being around 50\% of $\bar{l}_{11}$.
      Table~\ref{tab:horton.maxlvariation}
      shows comparisons for various river networks
      for both area and length data.
      }
    \label{fig:horton.lw_blownup_congo}
  \end{center}
\end{figure}

As noted in the previous
section, all moments of the generalized Horton distributions
grow exponentially with order.  
Coupling this with the fact that $n_\om \propto R_n^{-\om}$,
i.e., the number of samples of order $\om$ basins
decreases exponentially with $\om$,
we observe that a basin's $a$ and $l$ will potentially differ greatly
from values predicted by Horton's laws.  

To illustrate this, 
Figure~\ref{fig:horton.lw_blownup_congo} 
specifically shows the distributions
$P(l_{3})$ and $P(l_{4})$ scaled up to give $P(l_{11})$ for
the Congo river.
The actual Congo's length
measured at this 1000 meter resolution is
represented by the solid line and is around
57\% of the distribution's mean as indicated
by the dashed line.  
Nevertheless, we see that the measured length is within a standard
deviation of the predicted value.

\begin{table}[htb!]
  \begin{center}
    \ifthenelse{\boolean{@twocolumn}}
    {
    \begin{tabular}{l|ccccc} \hline\hline \\
      basin: &  $l_\Om$ & $\bar{l}_\Omega$ & 
      $\sigma_l$ & $l_\Om/\bar{l}_\Omega$ & $\sigma_l/\bar{l}_\Omega$ \\ \hline
      Mississippi & 4.92 & 11.10 & 5.60 & 0.44 & 0.51 \\
      Amazon &  5.75 & 9.18 & 6.85 & 0.63 & 0.75 \\
      Nile & 6.49 & 2.66 & 2.20 & 2.44 & 0.83 \\
      Congo & 5.07 & 10.13 & 5.75 & 0.50 & 0.57 \\
      Kansas & 1.07 & 2.37 & 1.74 & 0.45 & 0.73 \\ \hline
         & $a_\Om$ & $\bar{a}_\Omega$ & $\sigma_a$ & $a_\Om/\bar{a}_\Omega$ & $\sigma_a/\bar{a}_\Omega$ \\ \hline
      Mississippi &  2.74 & 7.55 & 5.58 & 0.36 & 0.74 \\
      Amazon & 5.40 & 9.07 & 8.04 & 0.60 & 0.89 \\
      Nile & 3.08 & 0.96 & 0.79 & 3.19 & 0.82 \\
      Congo & 3.70 & 10.09 & 8.28 & 0.37 & 0.82 \\
      Kansas & 0.14 & 0.49 & 0.42 & 0.28 & 0.86 \\ \hline\hline
    \end{tabular}
    }
    {
    \begin{tabular}{l|ccccc|ccccc}
      basin: &  $l_\Om$ & $\bar{l}_\Omega$ & 
      $\sigma_l$ & $l_\Om/\bar{l}_\Omega$ & $\sigma_l/\bar{l}_\Omega$ & $a_\Om$ & $\bar{a}_\Omega$ & $\sigma_a$ & $a_\Om/\bar{a}_\Omega$ & $\sigma_a/\bar{a}_\Omega$ \\ \hline
      Mississippi & 4.92 & 11.10 & 5.60 & 0.44 & 0.51 & 2.74 & 7.55 & 5.58 & 0.36 & 0.74 \\
      Amazon &  5.75 & 9.18 & 6.85 & 0.63 & 0.75 & 5.40 & 9.07 & 8.04 & 0.60 & 0.89 \\
      Nile & 6.49 & 2.66 & 2.20 & 2.44 & 0.83 & 3.08 & 0.96 & 0.79 & 3.19 & 0.82 \\
      Congo & 5.07 & 10.13 & 5.75 & 0.50 & 0.57 & 3.70 & 10.09 & 8.28 & 0.37 & 0.82 \\
      Kansas & 1.07 & 2.37 & 1.74 & 0.45 & 0.73 & 0.14 & 0.49 & 0.42 & 0.28 & 0.86 \\ \hline\hline
    \end{tabular}
    }
    \caption[Predicted versus measured main stream lengths]{
      Comparison of predicted versus measured main stream lengths
      for large scale river networks.  
      The dimensions of all lengths and areas
      are $10^6$ m and $10^{12}$ m$^2$ respectively.
      Here, $l_\Om$, is the actual main stream
      length of the basin, $\bar{l}_\Omega$ the
      predicted mean value of $l_\Om$, $\sigma_l$ 
      the predicted variance and $\sigma_l/\bar{l}_\Omega$
      the normalized deviation.
      The entries for the basin area data have
      corresponding definitions.
      }
    \label{tab:horton.maxlvariation}
  \end{center}
\end{table}

In Table~\ref{tab:horton.maxlvariation},
we provide a comparison of predicted 
versus measured main stream lengths
and areas for the basins studied here.
The mean for the scaled up distributions
overestimates the actual values in
all cases except for the Nile.
Also, apart from the Nile, all values
are within a standard deviation
of the predicted mean.
The coefficients of variation, $\sigma_a/\bar{a}_\Omega$
and $\sigma_l/\bar{l}_\Omega$, all indicate
that fluctuations are on the order of the expected
values of stream lengths and areas.

Thus, we see that by using a probabilistic point of
view, this generalized notion of Horton's laws
provides a way of discerning the strength
of deviations about the expected mean.
In general, stronger deviations would imply 
that geologic conditions play a more significant role
in setting the structure of the network.

\section{Conclusion}
The objective of this work has been to explore
the underlying distributions of river network
quantities defined with stream ordering.
We have shown that functional relationships
generalize all cases of Horton's laws.
We have identified
the basic forms of the distributions for
stream segment lengths (exponential)
and main stream lengths (convolutions
of exponentials) and shown a link
between number and area distributions.
Data from the
continent-scale networks of the Mississippi,
Amazon, and Nile river basins as well as from
Scheidegger's model of directed random networks 
provide both agreement with and inspiration
for the generalizations of Horton's laws.
Finally, we have identified a
fluctuation length scale $\xi$
which is a reinterpretation of what was previously
identified as only a mean value.
We see the study of
the generalized Horton distributions as
integral to increasing our 
understanding of river network structure.
We also suggest that practical network analysis 
be extended to measurements of distributions
and the length scale $\xi$ with the aim of
refining our ability to distinguish and 
compare network structure.

By taking account of fluctuations inherent in network
scaling laws, we are able to see how measuring Horton's
laws on low-order networks is unavoidably problematic.
Moreover, as we have observed, the measurement of the Horton
ratios is in general a delicate operation suggesting
that many previous measurements are not without error.

The theoretical understanding of the growth
and evolution of river networks requires
a more thorough approach to measurement
and a concurrent improvement in the statistical
description of river network geometry.
The present consideration of a generalization
of Horton's laws is a necessary step in this process
giving rise to stronger tests of both real and synthetic data.
In the following paper~\cite{dodds2000uc}, we round out this
expanded picture of network structure by
consdering the spatial distribution of network components.

\section*{Acknowledgements}
\label{sec:horton.ack}
The work was supported in part by NSF grant EAR-9706220
and the Department of Energy grant DE FG02-99ER 15004.
The authors would like to express their gratitude 
to H.\ Cheng for enlightening and enabling discussions.

\appendix

\section{Analytic connections between stream length distributions}
\label{sec:horton.theory}

In this appendix we consider a series of analytic calculations.
These concern the connections between the distributions
of stream segment lengths $\okellsom$, 
ordered basin main stream lengths $l_\om$
and main stream lengths $l$.  
We will idealize the problem
in places, assuming perfect scaling and infinite
networks while making an occasional salubrious approximation.
Also, we will treat the problem of lengths fully noting
that derivations of distributions for areas follow
similar but more complicated lines.

We begin by rescaling the form of 
stream segment length distributions
\begin{equation}
  \label{eq:horton.ellwfreq2}
  P(\okellsom,\om) = (R_n-1)(R_nR_\okell)^{-\om} F_\okell(lR_\okell^{-\om}).
\end{equation}
The normalization $c_\okell = R_n-1$ stems from the
requirement that
\begin{equation}
  \label{eq:horton.ellwfreq2a}
  \int_{u=0}^\infty F_\okell(u) = 1,
\end{equation}
which is made purely for aesthetic purposes.
As we have suggested in equation~\req{eq:horton.elldist}
and demonstrated empirical support for,
$F_\okell(u)$ is well approximated by the exponential
distribution $\xi^{-1} e^{-u/\xi}$.  For low $u$ and also
we have noted that deviations do of course
occur but they are sufficiently insubstantial as to
be negligible for a first order treatment of the problem.

\subsection{Distributions of main stream lengths as a function of stream order}
\label{subsec:horton.mainstreamdist}

We now derive a form for the distribution of 
main stream lengths $P(l_\om|\om)$.
As we have discussed, since $l_\om = \sum_{i=1}^{\om} \okellsom$, 
we have the convolution~\req{eq:horton.l-ellconv}.
The right-hand side of equation~\req{eq:horton.l-ellconv}
consists of exponentials 
as per equation~\req{eq:horton.elldist} so we now consider
the function $K_\om(u;\vec{a})$ given by
\begin{equation}
  \label{eq:horton.K_om-def}
  K_\om(u;\vec{a}) = 
  a_{1}e^{-a_{1}u} \ast a_{2}e^{-a_{2}u} \ast \cdots \ast a_{\om}e^{-a_{\om}u},
\end{equation}
where $\vec{a} = (a_1,a_2,\ldots,a_\om)$.
We are specifically interested in the case
when no two of the $a_i$ are equal, i.e.,
$a_i \neq a_j$ for all $i \neq j$.
To compute this $\om$-fold convolution, we simply examine the
$K_\om(u;\vec{a})$ for $\om=2$ and $\om=3$ and identify the emerging pattern.
For $\vec{a} = (a_1,a_2)$ we have,
omitting the prefactors for the time being,
\begin{eqnarray}
  \label{eq:horton.2expconv}
  \lefteqn{e^{-a_{1}u} \ast e^{-a_{2}u}} \nonumber \\
& = &\frac{e^{-a_{1}u}-e^{-a_{2}u}}{a_1-a_2}
    = \frac{e^{-a_{1}u}}{a_1-a_2}
    + \frac{e^{-a_{2}u}}{a_2-a_1}
\end{eqnarray}
providing $a_1 \neq a_2$.  Convolving this with $e^{-a_{3}u}$
we obtain
\begin{eqnarray}
  \label{eq:horton.3expconv}
  \lefteqn{e^{-a_{1}u} \ast e^{-a_{2}u} \ast e^{-a_{3}u} = 
    \left(\frac{e^{-a_{1}u}-e^{-a_{2}u}}{a_1-a_2}\right)
    \ast e^{-a_{3}u}}, \nonumber \\
  & = & \frac{e^{-a_{1}u}-e^{-a_{3}u}}{(a_1-a_2)(a_1-a_3)} -
  \frac{e^{-a_{2}u}-e^{-a_{3}u}}{(a_1-a_2)(a_2-a_3)}, \nonumber \\
  & = & \frac{e^{-a_{1}u}}{(a_1-a_2)(a_1-a_3)} + \nonumber \\
  &   & \frac{e^{-a_{2}u}}{(a_2-a_1)(a_2-a_3)} +
  \frac{e^{-a_{3}u}}{(a_3-a_1)(a_3-a_2)}.
\end{eqnarray}

Generalizing from this point, we obtain
\begin{equation}
  \label{eq:horton.K_om-calc}
  K_\om(u;\vec{a}) = 
  \left(\prod_{i=1}^{\om} a_i \right)
  \sum_{i=1}^{\om} \frac{e^{-a_{i}u}}
  {\prod_{j=1,j\neq i}^{\om}(a_i-a_j)}.
\end{equation}
Now, setting $a_i=1/(\xi (R_\okell)^{i-1})$ and carrying
out some manipulations we obtain the following expression
for $P(l_\om,\om)$:
\begin{widetext}
\begin{eqnarray}
  \label{eq:horton.Pl_om,omdist}
  \lefteqn{P(l_\om,\om) = 
    \frac{1}{(R_n)^\om} 
    \frac{1}{\prod_{j=1}^{\om}
      \xi (R_\okell)^{i-1}}
    \sum_{i=1}^{\om} \frac{e^{-l_\om/\xi (R_\okell)^{i-1}}}
    {\prod_{j=1,j\neq i}^{\om}(1/\xi(R_\okell)^{i-1}-1/\xi(R_\okell)^{j-1})}},
  \nonumber\\
  & = & \frac{1}{(R_n)^\om} 
  \frac{1}{\xi^\om \prod_{j=1}^{\om}(R_\okell)^{j-1}}
  \sum_{i=1}^{\om} e^{-l_\om/\xi (R_\okell)^{i-1}}
  \xi^{\om-1}
  \frac{\prod_{j=1,j\neq i}^{\om}(R_\okell)^{i-1}
    \prod_{j=1,j\neq i}^{\om}(R_\okell)^{j-1}}
  {\prod_{j=1,j\neq i}^{\om}(R_\okell)^{j-1} - (R_\okell)^{i-1}},
  \nonumber\\
  & = & \frac{1}{(R_n)^\om} 
  \frac{\xi^{\om-1}}{\xi^\om}
  \sum_{i=1}^{\om} e^{-l_\om/\xi (R_\okell)^{i-1}}
  \frac{(R_\okell)^{-2(i-1)} \prod_{j=1}^{\om}(R_\okell)^{i-1}
    \prod_{j=1}^{\om}(R_\okell)^{j-1}
    \prod_{k=1}^{\om}(R_\okell)^{-(j-1)} }
  {(R_\okell)^{-(\om-1)}\prod_{j=1,j\neq i}^{\om}(R_\okell)^{j} - (R_\okell)^{i}},
  \nonumber\\
  & = & \frac{1}{(R_n)^\om}
  \frac{1}{\xi}
  \sum_{i=1}^{\om} e^{-l_\om/\xi (R_\okell)^{i-1}}
  \frac{(R_\okell)^{(i-1)(\om-2)} (R_\okell)^{\om-2}/R_\okell}
  {\prod_{j=1,j\neq i}^{\om}(R_\okell)^{j} - (R_\okell)^{i}},
  \nonumber \\
  & = & \frac{1}{(R_n)^\om}
  \frac{1}{\xi R_\okell}
  \sum_{i=1}^{\om} e^{-l_\om/\xi (R_\okell)^{i-1}}
  \frac{(R_\okell)^{i(\om-2)}}
  {\prod_{j=1,j\neq i}^{\om}(R_\okell)^{j} - (R_\okell)^{i}} 
\end{eqnarray}
\end{widetext}
Note that we have added in a factor of
$1/(R_n)^\om$ for the appropriate normalization.
In addition, one observes that $P(0,\om)=0$ for
all $\om>1$ since all convolutions of pairs of
exponentials vanish at the origin.
Furthermore, the tail of the distribution is dominated by 
the exponential corresponding to the largest stream segment.

The next step is to connect to the power law distribution
of main stream lengths, $P(l)$
(see Figure~\ref{fig:lw_powerlawsum2_mispi}
and the accompanying discussion).
On considering equation~\req{eq:horton.l-lom}
we see that the problem can possibly be addressed with
some form of asymptotic analysis.

Before attacking this calculation however, 
we will simplify the notation keeping only the important
details of the $P(l_\om,\om)$.  Our main interest
is to see how equation~\req{eq:horton.l-lom} gives
rise to a power law.
We transform the outcome of equation~\req{eq:horton.Pl_om,omdist}
by using $n=\om$, $u=l_\om/\xi$, $r=R_\okell$, and $s=R_n$, neglecting
multiplicative constants and then summing
over stream orders to obtain
\begin{equation}
  \label{eq:horton.nexpconvgeomratio2}
  G(u) = 
  \sum_{n=1}^{\infty} \frac{1}{s^{n}}
  \sum_{i=1}^{n} 
  \frac{r^{(n-2)i} e^{-u/r^{i-1}}}
  {\prod_{j=1, j \neq i}^{n}(r^j-r^i)}.
\end{equation}
The integration over $l_\om$ has been omitted
meaning that the result will be a power law
with one power lower than expected.

\subsection{Power law distributions of main stream lengths}
\label{subsec:horton.powerlawdists}

We now show that this sum of exponentials $G(u)$
in equation~\req{eq:horton.nexpconvgeomratio2}
does in fact asymptotically tend to a power law.
We first interchange the order of summation
replacing $\sum_{n=1}^{\infty}\sum_{i=1}^{n}$
with $\sum_{i=1}^{\infty}\sum_{n=i}^{n}$
to give
\begin{eqnarray}
  \label{eq:horton.nexpconvgeomratio2_mod2}
  G(u) & = &
  \sum_{i=1}^{\infty}
  e^{-u/r^{i-1}} 
  \sum_{n=1}^{\infty} 
  \frac{r^{(n-2)i}}
  {s^{n}\prod_{j=1, j \neq i}^{n}(r^j-r^i)},
  \nonumber \\
  & = & \sum_{i=1}^\infty C_i   e^{-u/r^{i-1}}.
\end{eqnarray}
We thus simply have a sum of exponentials to contend with.
The coefficients $C_i$ appear unwieldy at first but
do yield a simple expression after some algebra which
we now perform:
\begin{eqnarray}
  \label{eq:horton.Ci}
\lefteqn{  C_i = \sum_{n=1}^{\infty} 
  \frac{r^{(n-2)i}}
  {s^{n}\prod_{j=1, j \neq i}^{n}(r^j-r^i)},}
  \nonumber \\
  & = & \frac{1}{\prod_{j=1}^{i-1}(r^j-r^i)}
  \sum_{n=i}^{\infty} 
  \frac{r^{(n-2)i}}
  {s^{n}\prod_{j=i+1}^{n}(r^j-r^i)},
  \nonumber \\
  & = &
    \frac{r^{(i-2)i}}{\prod_{j=1}^{i-1}(r^j-r^i)}
  \frac{1}{s^{i}}
  \sum_{n=i}^{\infty} 
  \frac{s^{i} r^{(n-2)i} r^{-(i-2)i}}
  {s^{n}\prod_{j=i+1}^{n}(r^j-r^i)},
  \nonumber \\
  & = & 
    \frac{1}{\prod_{j=1}^{i-1}r^{-i}(r^j-r^i)}
    \frac{r^{-i}}{s^{i}}
  \sum_{n=i}^{\infty} 
  \frac{r^{(n-i)i}}{s^{n-i}\prod_{j=i+1}^{n}(r^j-r^i)},
  \nonumber \\
  & = & 
    \frac{1}{\prod_{j=1}^{i-1}(r^{j-i}-1)}
    \frac{1}{r^{i}s^{i}}
  \sum_{n=i}^{\infty} 
  \frac{1}{\prod_{j=i+1}^{n}sr^{-i}(r^j-r^i)},
  \nonumber \\
  & = & \frac{1}{r^{i}s^{i}} 
    \frac{1}{\prod_{j=1}^{i-1}(r^{j-i}-1)}
  \sum_{n=i}^{\infty} 
  \prod_{j=i+1}^{n} \frac{1}{s(r^{j-i}-1)},
  \nonumber \\
  & = & \frac{1}{r^{i}s^{i}}
  \frac{1}{\prod_{j=1}^{i-1}(r^{j-i}-1)}
  \sum_{n=i}^{\infty} 
  \prod_{j=i+1}^{n} \frac{1}{s(r^{j-i}-1)},
  \nonumber \\
  & = & \frac{1}{r^{i}s^{i}}
  \left(\frac{-1}{\prod_{k=1}^{i-1}(1-r^{-k})}
  \right)
  \left(
  \sum_{m=1}^{\infty} 
  \prod_{k=1}^{m} \frac{1}{s(r^{k}-1)}
  \right).
\end{eqnarray}
In reaching the last line we have shifted the indices
in several places.
In the last bracketed term we have set $k=j-i$ and then $m=n-i$
while in the first bracketed term, we have used $-k=j-i$.
Immediately of note is that the last term is independent
of $i$ and may thus be ignored.  

The first bracketed term
does depend on $i$ but converges rapidly.
Writing $D_i = \prod_{k=1}^{i-1}(1-r^{-k})$
we have that $D_i = D_m\prod_{k=m}^{i-1}(1-r^{-k})$.
Taking $m$ to be fixed and large enough such
that $1-r^{-k}$ is approximated well by $\exp\{-r^{-k}\}$
for $k \ge m$, we then have
\begin{eqnarray}
  \label{eq:horton.DiDm}
  D_i & = & D_m \exp\left\{ \sum_{k=m}^{i-1}-r^{-k} \right\},
  \nonumber \\
  & = & D_m \exp\left\{ \frac{r^{1-m}}{(r-1)} (1-1/r^{i-m-1})\right\}.
\end{eqnarray}
As $i\rightarrow\infty$, $D_i$ clearly approaches a product
of $D_m$ and a constant.
Therefore, the first bracketed term in equation~\req{eq:horton.Ci}
may also be neglected in an asymptotic analysis.

Hence, as $i \rightarrow \infty$,
the coefficients $C_i$ are simply given by
\begin{equation}
  \label{eq:horton.Ci_2}
  C_i \propto \frac{1}{s^i r_i}.
\end{equation}
and we can approximate $G(u)$ as, boldly using the
equality sign,
\begin{equation}
  \label{eq:horton.nexpconvgeomratio2_mod3}
  G(u) = AS(u) = A\sum_{i=0}^\infty \frac{e^{-u/r^{i}}}{r^{i(1+\gamma)}},
\end{equation}
where $A$ comprises the constant part of the $C_i$
and factors picked up by shifting the lower limit of the
index $i$ from $1$ to $0$.  
We have also used here the identification
\begin{equation}
  \label{eq:horton.srgamma}
  s = r^\gamma.
\end{equation}
We turn now to the
asymptotic behavior of $S(u)$,
this being the final stretch of our analysis

There are several directions one may take at this point.
We will proceed by employing a transformation of $S(u)$ that
is sometimes referred to as the Sommerfeld-Watson
transformation and also as Watson's lemma~\citep[p.~239]{carrier66}.
Given a sum over any set of integers $I$, say 
$S = \sum_{n \in I} f(n)$, it can be written as the
following integral
\begin{equation}
S = \frac{1}{2 \pi i}\oint_C 
\frac{\pi\cos{\pi z}}{\sin{\pi z}} f(z) dz.
\label{eq:sommerfeld-watson-tf}
\end{equation}
where $C$ is a contour that contains the points on
the real axis $n + i0$ where $n \in I$ and none
of the points of the same form with $n \in \mathbb{Z}/I$.
Calculation of the residues of the simple poles of the 
integrand return us to the original sum.

\begin{figure}[htb!]
  \begin{center}
    \epsfig{file=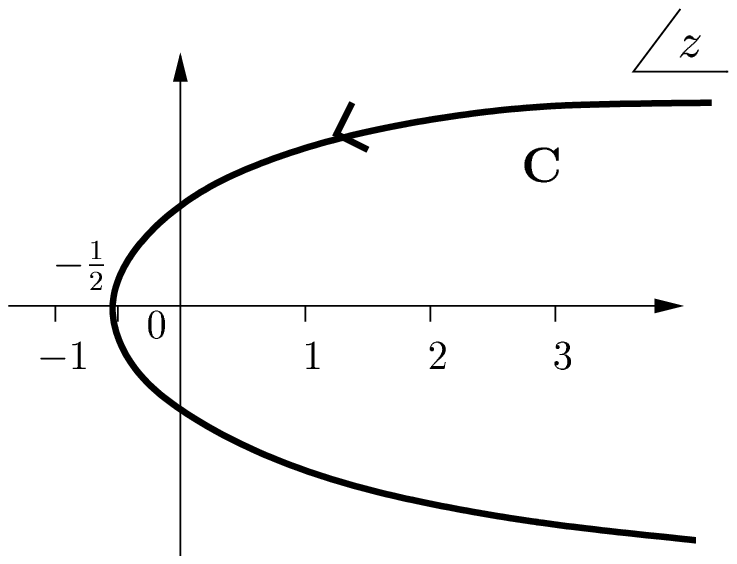,width=0.4\textwidth}
    \caption[Integration contour for equation~\req{eq:horton.S(u)-transform}]{
      Contour $C$ used for evaluation of the integral
      given in equation~\req{eq:horton.S(u)-transform}.  
      The poles are situated
      at $n+0i$ where $n \in \{0,1,2,\ldots\}$.
      }
    \label{fig:horton.contour-swtf}
  \end{center}
\end{figure}

Applying the transformation to $S(u)$ we obtain
\begin{equation}
  \label{eq:horton.S(u)-transform}
  S(u) = \frac{1}{2 \pi i}\oint_C 
  \frac{\pi\cos{\pi z}}{\sin{\pi z}} 
  e^{-u r^{-z}} r^{-z(1+\gamma)} \dee{z}.
\end{equation}
The contour $C$ is represented in 
Figure~\ref{fig:horton.contour-swtf}.

\begin{figure}[htb!]
  \begin{center}
    \epsfig{file=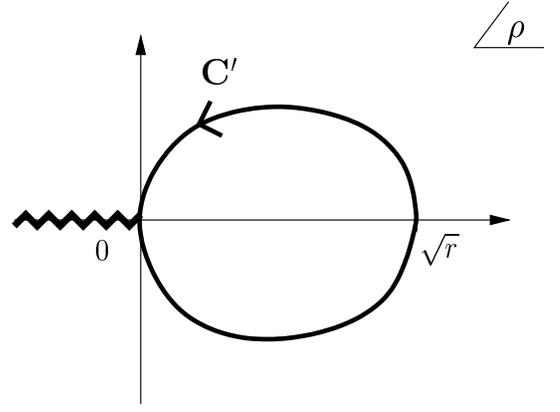,width=0.4\textwidth}
    \caption[Integration contour for equation~\req{eq:horton.S(u)-transform2}]{
      Contour $C'$ used for evaluation of the integral
      given in equation~\req{eq:horton.S(u)-transform2}
      as deduced from contour $C$ (Figure~\ref{fig:horton.contour-swtf})
      with the transformation $\rho=r^{-z}$.  
      The negative real axis is a branch cut.
      }
    \label{fig:horton.contour-swtf2}
  \end{center}
\end{figure}

We first make a change of variables, $r^{-z} = \rho$.
Substituting this and $\dee{z} = -\dee{\rho}/\rho\ln{r}$
into equation~\req{eq:horton.S(u)-transform} we have
\begin{eqnarray}
  \label{eq:horton.S(u)-transform2}
  S(u) & = &\frac{1}{2 \pi i}
  \oint_C'
  \frac{\pi\cos{-\pi\ln{\rho}/\ln{r}}}{\sin{-\pi\ln{\rho}/\ln{r}}}
  e^{-u\rho} \rho^{(1+\gamma)} (-\dee{\rho}/\rho\ln{r}) 
  \nonumber \\
  & = & \frac{1}{2 i \ln r}
  \oint_C'
  \frac{\pi\cos{\pi\ln{\rho}/\ln{r}}}{\sin{\pi\ln{\rho}/\ln{r}}}
  e^{-u\rho} \rho^{\gamma} \dee{\rho}.
\end{eqnarray}
The transformed contour $C'$ is depicted in
Figure~\ref{fig:horton.contour-swtf2}.

\begin{figure}[htb!]
  \begin{center}
    \epsfig{file=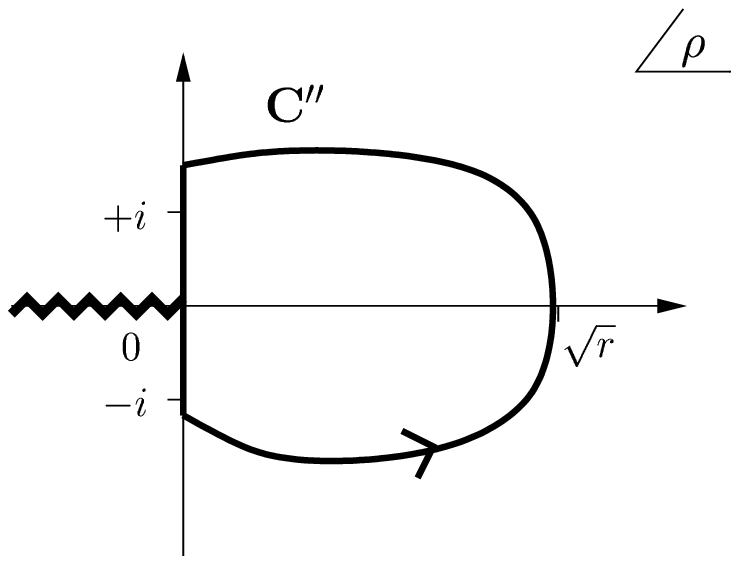,width=0.4\textwidth}
    \caption[Integration for equation~\req{eq:horton.S(u)-transform}]{
      Contour $C''$ used for evaluation of the integral
      given in equation~\req{eq:horton.S(u)-transform}.  
      The poles are situated 
      at $n+0i$ where $n \in \{0,1,2,\ldots\}$.
      }
    \label{fig:horton.contour-swtf3}
  \end{center}
\end{figure}

As $u\rightarrow\infty$, the contribution to integral from 
the neighborhood of $\rho=0$ dominates.  The introduction
of the sin and cos terms has created an interesting oscillation
that has to be handled with with some care.  We now deform
the integration contour $C'$ into the contour 
$C''$ of Figure~\ref{fig:horton.contour-swtf3}
focusing on the interval along the imaginary axis $[-i,i]$.
Choosing this path will simplify the cos and sin expressions
which at present have logs in their arguments.

The integral $S(u)$ is now given by
$S(u) \simeq I(u) + \mbox{c.c.}$ where
\begin{equation}
  \label{eq:horton.S(u)-transform3}
  I(u) = \frac{-1}{2i\ln{r}}
  \int_0^{i}
  \frac{\pi\cos{\pi\ln{\rho}/\ln{r}}}{\sin{\pi\ln{\rho}/\ln{r}}}
  e^{-u\rho} \rho^{\gamma} \dee{\rho}.
\end{equation}
Writing $\rho = \sigma + i\tau$ with $\sigma=0$, 
we have $\dee{\rho} = i\dee{\tau}$
and the following for the cos and sin terms:
\begin{eqnarray}
  \label{eq:horton.cos}
\lefteqn{  \cos{\pi\ln{\rho}/\ln{r}} = 
  \frac{\rho^{i\pi/\ln{r}} + \rho^{-i\pi/\ln{r}}}{2},} \nonumber \\
  & = & \frac{\tau^{i\pi/\ln{r}}e^{-\pi^2/2\ln{r}} 
    + \tau^{-i\pi/\ln{r}}e^{\pi^2/2\ln{r}}}{2},
\end{eqnarray}
and
\begin{eqnarray}
  \label{eq:horton.sin}
  \lefteqn{ \sin{\pi\ln{\rho}/\ln{r}}  = 
  \frac{\rho^{i\pi/\ln{r}} - \rho^{-i\pi/\ln{r}}}{2i},} \nonumber \\
  & = & \frac{\tau^{i\pi/\ln{r}}e^{-\pi^2/2\ln{r}} 
    - \tau^{-i\pi/\ln{r}}e^{\pi^2/2\ln{r}}}{2i}.
\end{eqnarray}
The $\cot$ term in the integrand becomes
\begin{eqnarray}
  \label{eq:horton.cot}
  \frac{\cos{\pi\ln{\rho}/\ln{r}}}{\sin{\pi\ln{\rho}/\ln{r}}}
  & = & -i\frac{1+\tau^{2i\pi/\ln{r}}e^{-\pi^2/\ln{r}}}
  {1-\tau^{2i\pi/\ln{r}}e^{-\pi^2/\ln{r}}} \nonumber \\
  & = & -i\frac{1+\delta(\tau)}{1-\delta(\tau)},
\end{eqnarray}
where $\delta(\tau)=\tau^{2i\pi/\ln{r}}e^{-\pi^2/\ln{r}}$.
The integral $I(u)$ now becomes
\begin{eqnarray}
  \label{eq:horton.Irho_2_tau}
  I(u) & = &
  \frac{i}{2\ln{r}}
  \int_{0}^{1}
  \frac{1+\delta(\tau)}{1-\delta(\tau)} e^{-iu\tau}
  \tau^{\gamma} e^{i\pi\gamma/2} \dee{\tau} 
  \nonumber \\
  & = &\frac{e^{i\pi(1+\gamma)/2}}{2\ln{r}}
  \int_{0}^{1}e^{-iu\tau}\tau^\gamma
  \frac{1+\delta(\tau)}{1-\delta(\tau)} \dee{\tau}.
\end{eqnarray}
Now, since 
$|\delta(\tau)| = e^{-\pi^2/\ln{r}} \lesssim 10^{-4}$
(taking $r=R_\okell\approx 2.5$), we can expand the
expression as follows
\begin{eqnarray}
  \label{eq:horton.deltaexpanse}
  \frac{1+\delta}{1-\delta}
   & = & (1+\delta)(1+\delta+\delta^2+\ldots) 
   \nonumber \\
   & = & 1 + 2\delta + 2\delta^2 + 2\delta^3 + \ldots
\end{eqnarray}
The integral in turn becomes 
\begin{eqnarray}
  \label{eq:horton.Iexpansion}
  \lefteqn{ I(u) = \frac{i^{1+\gamma}}{2\ln{r}}
  \int_{0}^{1} \dee{\tau} \tau^{\gamma} e^{-iu\tau} \times } \nonumber \\
  & & \left(
    1 + 2\tau^{2i\pi/\ln{r}}e^{-\pi^2/\ln{r}}
    + 2\tau^{4i\pi/\ln{r}}e^{-2\pi^2/\ln{r}}  + \ldots \right. \nonumber \\
  & & \left.
    \mbox{} + 2\tau^{2ni\pi/\ln{r}}e^{-n\pi^2/\ln{r}} + \ldots 
  \right) 
\end{eqnarray}

The basic $n$-th integral in this expansion is
\begin{equation}
  \label{eq:horton.Ipiece}
  I_n(u) = \int_{0}^{1} \tau^{\gamma+2ni\pi/\ln{r}} e^{-iu\tau} \dee{\tau}.
\end{equation}
Substituting $u\tau = w$ and replacing the upper limit
$w=u$ with $w=\infty$ we have 
\begin{eqnarray}
  \label{eq:horton.Ipiece2}
  \lefteqn{I_n(u)  = u^{-(1+\gamma+2ni\pi/\ln{r})}
  \int_{0}^{\infty} \dee{w} w^{\gamma+2ni\pi/\ln{r}} e^{-iw}, }
  \nonumber \\
  & = & (iu)^{-(1+\gamma+2ni\pi/\ln{r})}
  \int_{0}^{\infty} i\dee{w} (iw)^{\gamma+2ni\pi/\ln{r}} e^{-iw}, 
  \nonumber \\
  & = & (iu)^{-(1+\gamma+2ni\pi/\ln{r})}
  \int_{0}^{\infty} \dee{v} (v)^{\gamma+2ni\pi/\ln{r}} e^{-v},
  \nonumber \\
  & = & (iu)^{-(1+\gamma+2ni\pi/\ln{r})}
  \Gamma(\gamma+2ni\pi/\ln{r}).
\end{eqnarray}
Here, we have rotated the contour along the imaginary $iw$-axis
to the real $v$-axis and identified the integral with
the gamma function $\Gamma$~\cite{gradshteyn65}.
The integral can now be expressed as
\begin{equation}
  \label{eq:horton.IGamma}
  I(u) = \frac{1}{2\ln{r}u^{1+\gamma}}
  \left[ 1 + 2\sum_{n=1}^{\infty} u^{-2ni\pi/\ln{r}}
    \Gamma(\gamma+2ni\pi/\ln{r}) \right].
\end{equation}
We now need to show that the higher order terms are
negligible.  Note that their magnitudes do no vanish with
increasing $u$ but instead are highly oscillatory terms.
Using the asymptotic form of the Gamma function~\cite{bender78}
\begin{equation}
  \label{eq:horton.gamma_asymptotic}
  \Gamma(z) = z^{z-1/2} e^{-z} \sqrt{2\pi}
  \left(1 + O(1/z) \right),
\end{equation}
we can estimate as follows for large $n$ that
\begin{eqnarray}
  \label{eq:horton.gamma_est}
  \lefteqn {|\Gamma(1+\gamma+2ni\pi/\ln{r})| } \nonumber \\
  & \sim & |(2i\pi n/\ln{r}+1+\gamma)^{2i\pi n/\ln{r} + 1/2 + \gamma}
  e^{-\gamma-1} \sqrt{2\pi} | 
  \nonumber \\
  & = & |(e^{i\pi/2} 2\pi n/\ln{r})^{2i\pi n/\ln{r} + 1/2 + \gamma}
  e^{-\gamma-1} \sqrt{2\pi} |
  \nonumber \\
  & = & e^{-\pi^2 n/\ln{r}} n^{\gamma+1/2}
  (2\pi/e)^{1+\gamma} (\ln{r})^{-1/2-\gamma}.
\end{eqnarray}
Hence, $\Gamma(1+\gamma+2ni\pi/\ln{r})$ vanishes
exponentially with $n$.
For the first few values of $n$ taking 
$\gamma=3/2$ and $r=2.5$, we 
have $\Gamma(1+\gamma+2i\pi/\ln{r}) \simeq 2.5\times{10^{-3}}$
and $\Gamma(1+\gamma+4i\pi/\ln{r}) \simeq 2.1\times{10^{-6}}$
showing that these corrections are negligible.

Hence we are able estimate 
$S(u)$ to first order as
\begin{equation}
  \label{eq:horton.IGamma2}
  S(u) \simeq \frac{1}{\ln{r}} u^{-1-\gamma}.
\end{equation}
Thus we have determined that a power law
follows from the initial assumption
that stream segment lengths follow
exponential distributions.

This equivalence has been drawn as
an asymptotic one, albeit one where convergences
have been shown to be rapid.
The calculation is clearly not the entire picture
as the solution does contain small rapidly-oscillating
corrections that do not vanish with increasing argument.
A possible remaining problem and
one for further investigation is to understand how
the distributions for main stream
lengths $l_\om$ fit together over a range that
is not to be considered asymptotic.
Nevertheless, the preceding is one attempt
at demonstrating this rather intriguing breakup
of a smooth power law into a discrete family
of functions built up from one fundamental
scaling function.


\begin{thebibliography}{10}
\bibitem{note:horton.niletable}
\bibinfo{note}{The data for the Nile comes from the United States Geological
  Survey's 30 arc second Hydro1K dataset, avaliable on the Internet at
  \www{edcftp.cr.usgs.gov}, which has a grid spacing of approximately 1000
  meters. At this resolution, the Nile is an order $\Om=10$ basin.}
\end{thebibliography}
\end{document}